\documentclass[fleqn,usenatbib,useAMS]{mnras}

\usepackage{graphicx}	
\usepackage{amsmath}	
\usepackage{amssymb}	
\usepackage{multicol}        
\usepackage{bm}		
\usepackage{pdflscape}	
\usepackage{graphicx}	
\usepackage{natbib}
\usepackage{hyperref}
\usepackage[usenames, dvipsnames]{color}

\newcommand\rah{\mbox{$^{\mathrm h}$}}%
\newcommand\ram{\mbox{$^{\mathrm m}$}}%
\newcommand\ras{\mbox{$^{\mathrm s}$}}%

\newcommand\arcdeg{\mbox{$^\circ$}}%

\usepackage[T1]{fontenc}
\usepackage{ae,aecompl}

\usepackage{newtxtext,newtxmath}

\title[An all-sky survey of circular polarisation at 200 MHz]{An all-sky survey of circular polarisation at 200 MHz}

\author[Lenc et al.]{\parbox[t]{\textwidth}{
Emil~Lenc,$^{1,2,3}$\thanks{E-mail: emil.lenc@sydney.edu.au}
Tara~Murphy,$^{1,2}$
C.~R. Lynch,$^{1,2}$
D.~L. Kaplan,$^{4}$ 
S.~N. Zhang,$^{5}$
}
\\
$^1$ Sydney Institute for Astronomy, School of Physics, The University of Sydney, NSW 2006, Australia\\
$^2$ ARC Centre of Excellence for All-sky Astrophysics (CAASTRO)\\
$^3$ CSIRO Astronomy and Space Science, PO Box 76, Epping, NSW 1710, Australia\\
$^4$ Department of Physics, University of Wisconsin--Milwaukee, Milwaukee, WI 53201, USA\\
$^5$ School of Astronomy and Space Science, Nanjing University, 163 Xianlin Avenue, Nanjing 210023, China
}

\date{Accepted XXX. Received YYY; in original form ZZZ}

\pubyear{2016}

\begin{document}
\label{firstpage}
\pagerange{\pageref{firstpage}--\pageref{lastpage}}
\maketitle

\begin{abstract}
We present results from the first all-sky radio survey in circular polarisation. The survey uses the Murchison Widefield Array (MWA) to cover 30\,900 sq. deg., over declinations south of $+30\arcdeg$ and north of $-86\arcdeg$ centred at 200\,MHz (over a $169-231$\,MHz band). We achieve a spatial resolution of $\sim$$3\arcmin$ and a typical sensitivity of $3.0$\,mJy\,PSF$^{-1}$ over most of the survey region. We demonstrate a new leakage mitigation technique that reduces the leakage from total intensity into circular polarisation by an order of magnitude. In a blind survey of the imaged region, we detect 14 pulsars in circular polarisation above a $6\sigma$ threshold. We also detect six transient sources associated with artificial satellites. A targeted survey of 2\,376 pulsars within the surveyed region yielded 33 detections above $4\sigma$. Looking specifically at pulsars previously detected at 200 MHz in total intensity, this represents a $35\%$ detection rate. We also conducted a targeted survey of 2\,400 known flare stars, this resulted in two tentative detections above $4\sigma$. A similar targeted search for 1\,506 known exoplanets in the field yielded no detections above $4\sigma$. The success of the survey suggests that similar surveys at longer wavelength bands and of deeper fields are warranted.

\end{abstract}




\section{Introduction}

The vast majority of all-sky radio surveys to date have focused on sources emitting in total intensity (Stokes I) e.g. Westerbork Northern Sky Survey \citep[WENSS;][]{Rengelink:1997}, Sydney University Molonglo Sky Survey \citep[SUMSS;][]{Mauch:2003}, Galactic and Extragalactic All-Sky MWA Survey \citep[GLEAM;][]{Hurley-Walker:2017b} and the GMRT 150 MHz All-sky Radio Survey First Alternative Data Release \citep[TGSS ADR1;][]{Intema:2017}. The exception is the NRAO VLA Sky Survey \citep[NVSS;][]{Condon:1998} which also considered linear (Stokes Q and Stokes U) polarisation \citep{Taylor:2009v702p1230}. However, to date, there have been no all-sky radio surveys in circular polarisation (Stokes V).

Numerous astrophysical sources are known emit circular polarisation with relatively high degrees of fractional polarisation ($\geq 1\%$). These include pulsars \citep{You:2006,Noutsos:2015,Johnston:2017}, flare stars \citep{Lynch:2017b}, and Jupiter \citep{Seaquist:1969}. It is also anticipated that some exoplanets may also emit circular polarisation \citep{Winglee:1986,Zarka:2001,Murphy:2015,Lynch:2017a}. Weak levels of circular polarisation ($0.01-1\%$) have been observed in active galactic nuclei \citep[AGN;][]{Komesaroff:1984, Weiler:1983, Rayner:2000, Aller:2012}, intra-day variable sources \citep{Macquart:2000}, and are also expected in diffuse Galactic synchrotron emission \citep{Ensslin:2017} at fractions of less than $0.1\%$.

Observations of circular polarisation can inform us about the physical processes within these sources and propagation effects along the line of sight. Coherent emission processes can generate highly circularly polarised emission \citep{Macquart:2002, Macquart:2003}, whereas circular polarisation resulting from synchrotron radiation is generally very weak. Propagation effects can also cause circular polarisation through scintillation in a magnetised plasma, refraction effects near black holes, and linearly polarised radiation passing through a relativistic plasma \citep{Macquart:2002, Macquart:2003}. To confirm the source of circularly polarised emission generally requires a combination of detailed observations and theoretical modelling \citep[e.g.][]{OSullivan:2013}.

Compared to total intensity, only a small fraction of sources emit circularly polarised radiation resulting in a lower classical confusion limit. As such, for instruments that are confusion limited in total intensity, such as the Murchison Widefield Array \citep[MWA,][]{Tingay:2013}, greater sensitivity can be achieved when observing in circular polarisation. This is particularly beneficial for sources that have high degrees of fractional circular polarisation, however, the gain in sensitivity diminishes for sources that exhibit low degrees of fractional circular polarisation. A positive aspect of this is that most sources are generally weak in circular polarisation and so do not greatly contribute to side-lobe confusion. Therefore deconvolution is unnecessary, greatly simplifying processing.

Continuum observations in circular polarisation may also aid in the detection of pulsars missed using conventional means. For example, pulsars with complex orbits, sub-millisecond pulsars, and pulsars exhibiting significant pulse broadening \citep{Bhat:2004, Xue:2017, Geyer:2017}. Traditional search methods assume regular, well-separated pulses, but accelerations in compact orbits lead to significant computational difficulties and broadening can cause individual pulses to blend. Despite this, if the pulsar is sufficiently steep-spectrum low-frequency imaging searches \citep[e.g.,][]{Frail:2018} may discover a number of pulsars that would be otherwise missed.  Even this can be problematic, though, as the noise in total intensity images can be significantly higher in the Galactic plane where most pulsars are found.  Therefore searches done in circular polarisation may allow the deepest searches for continuum emission independent of other pulsar properties. Searching in circular polarisation is further beneficial compared to searches in Stokes I continuum as very few sources exhibit circularly polarised emission and so the number of candidate sources greatly reduces. Transient searches are also simplified as they are less affected by source confusion \citep[e.g.][]{Lynch:2017b}.

Despite the potential for discovery available with observations in circular polarisation, there have been no all-sky surveys to date. All observations have been targeted towards specific known source populations e.g. AGN \citep{Rayner:2000, Cenacchi:2011}, pulsars \citep{Johnston:2017} and exoplanets \citep{Murphy:2015}. For the most part, all-sky observations of circular polarisation have been hindered by instrumental leakage. In the case of dipole-based instruments, such as the MWA, this leakage is primarily caused by poor models of the primary beam \citep{Sutinjo:2015v50p52S}. For the MWA, the effect is particularly pronounced at higher frequencies and towards the edge of the beam. \citet{Lenc:2017} demonstrated that polarisation leakage observed with the MWA could be mitigated in drift-scan observations by modelling the leakage pattern across the beam and then subtracting it.

In this paper, we present the first all-sky radio survey in circular polarisation. The survey covers the entire Southern sky ranging in declination from $-85\arcdeg$ to $+30\arcdeg$ at a frequency of 200\,MHz.  We use data originally observed as part  of the GLEAM survey \citep{Wayth:2015v32p25,Hurley-Walker:2017A}. In the process of performing this survey, we also tested the effectiveness of leakage mitigation. Throughout this paper we have adopted the PSR/IEEE convention for left-handed and right-handed circular polarisation \citep{vanStraten:2010} which are of positive and negative sign, respectively.

\section{Observations and data analysis}

\subsection{Observations}

We used archival visibility data observed as part of GLEAM \citep{Wayth:2015v32p25,Hurley-Walker:2017A}. The observations used a drift-scan observing mode where tiles always point at the meridian. As such, instrumental systematics are minimised by maintaining a consistent observing set up. To allow direct comparison to the GLEAM deep wide-band survey data \citep{Hurley-Walker:2017A} we used the $169-200$\,MHz and $200-231$\,MHz frequency bands. Inaccuracies in the MWA beam model make these two frequency bands more prone to polarisation leakage than the three lower bands \citep{Sutinjo:2015v50p52S,Lenc:2016,Sokolowski:2017}, so this data set is well-suited to testing the effectiveness of our polarisation leakage subtraction technique. A list of the individual GLEAM drift-scan observations used is summarised in Table \ref{table:obsdata}.

\subsection{Data reduction}

We calibrated the data using the real-time calibration and imaging system, referred to as the \textsc{rts} \citep{Mitchell:2008v2p707,Ord:2010}, using the procedure outlined in \citet{Lenc:2016} for point source polarimetry. Calibration was performed per epoch using a calibrator observation for that specific epoch (see Table \ref{table:obsdata}), apart from the 2013-08-18 epoch where the calibration solution for that day was poor and so a solution from the previous day was used. 

Archival online flagging \citep{Offringa:2012} was applied to reduce the effects of radio frequency interference. To minimise sidelobe confusion and reduce sensitivity to large-scale structure, robust weighting was used with a robustness of $-1$ and only baselines longer than $50\lambda$ were utilised. Dirty image cubes were created for each two-minute snapshot using the \textsc{rts} at full 40\,kHz spectral resolution and over a $25\arcdeg\times25\arcdeg$ region centred on the beam pointing location. All images were $2\,187\times2\,187$ pixels in extent and with a pixel size of $\sim$$41\arcsec$, this equates to a sampling of $\sim$$5$ pixels across the beam at $200\,$MHz. The spectral cubes were averaged in frequency to create two-minute snapshot images for Stokes I, Q, U and V.

\begin{table}
\centering
\begin{tabular}{l r l r l }
\hline \hline
Date       & Dec.           & RA range (h)& $N_\text{flag}$  & Calibrator \\ 
\hline
2013-08-08 & $+1.6\arcdeg$  & $19.5-3.5$ & 3     & 3C444 \\
2013-08-09 & $-55.0\arcdeg$ & $19.5-3.5$ & 11    & 3C444 \\
2013-08-10 & $-26.7\arcdeg$ & $19.5-3.5$ & 11    & 3C444 \\
2013-08-17 & $+18.6\arcdeg$ & $19.5-3.5$ & 4     & 3C444 \\
2013-08-18 & $-72.0\arcdeg$ & $19.5-3.5$ & 6     & 3C444\textsuperscript{a} \\
2013-08-22 & $-13.0\arcdeg$ & $19.5-3.5$ & 1     & 3C444 \\
2013-08-25 & $-40.0\arcdeg$ & $19.5-3.5$ & 1     & 3C444 \\
2013-11-05 & $-13.0\arcdeg$ & $0-8$      & 5     & Hydra A \\
2013-11-06 & $-40.0\arcdeg$ & $0-8$      & 6     & Hydra A \\
2013-11-07 & $+1.6\arcdeg$  & $0-8$      & 4     & Hydra A \\
2013-11-08 & $-55.0\arcdeg$ & $0-8$      & 6     & Hydra A \\
2013-11-11 & $+18.6\arcdeg$ & $0-8$      & 8     & Hydra A \\
2013-11-12 & $-72.0\arcdeg$ & $0-8$      & 18    & Hydra A \\
2013-11-25 & $-26.7\arcdeg$ & $0-8$      & 0     & Hydra A \\
2014-03-03 & $-26.7\arcdeg$ & $6-16$     & 0     & Hydra A \\
2014-03-04 & $-13.0\arcdeg$ & $6-16$     & 0     & Hydra A \\
2014-03-06 & $+1.6\arcdeg$  & $6-16$     & 1     & Hydra A \\
2014-03-08 & $+18.6\arcdeg$ & $6-16$     & 1     & Hydra A \\
2014-03-09 & $-72.0\arcdeg$ & $6-16$     & 1     & Hydra A \\
2014-03-16 & $-40.0\arcdeg$ & $6-16$     & 1     & Hydra A \\
2014-03-17 & $-55.0\arcdeg$ & $6-16$     & 1     & Hydra A \\
2014-06-09 & $-26.7\arcdeg$ & $12-22$    & 3     & 3C444 \\
2014-06-10 & $-40.0\arcdeg$ & $12-22$    & 4     & 3C444 \\
2014-06-11 & $+1.6\arcdeg$  & $12-22$    & 5     & 3C444 \\
2014-06-12 & $-55.0\arcdeg$ & $12-18.5$  & 6     & Hercules A \\
2014-06-13 & $-13.0\arcdeg$ & $12-19$    & 5     & Hercules A \\
2014-06-14 & $-72.0\arcdeg$ & $12-22$    & 4     & 3C444 \\
2014-06-15 & $+18.6\arcdeg$ & $12-22$    & 5     & 3C444 \\
2014-06-16\textsuperscript{b} & $-13.0\arcdeg$ & $18.5-22$  & 3     & 3C444 \\
2014-06-18\textsuperscript{c} & $-55.0\arcdeg$ & $15-22$    & 0     & 3C444 \\ [1ex]
\hline
\multicolumn{5}{l}{\textsuperscript{a}\footnotesize{Calibration from previous day (2013-08-17) re-used.}} \\
\multicolumn{5}{l}{\textsuperscript{b}\footnotesize{Partial reobservation of 2014-06-13 drift scan.}} \\
\multicolumn{5}{l}{\textsuperscript{c}\footnotesize{Partial reobservation of 2014-06-12 drift scan.}} \\
\end{tabular}
\caption{GLEAM first year observing parameters. $N_\text{flag}$ is the number of MWA tiles (of the 128 available) that are flagged. The calibrator is used to determine initial bandpass, phase and flux density scale corrections.}
\label{table:obsdata}
\end{table}

\subsection{Beam modelling and leakage}

The true beam pattern of the MWA, as measured empirically by imaging field sources, differs significantly from the analytic beam pattern at the higher end of the MWA band and/or when observing at lower elevation \citep{Sutinjo:2015v50p52S}. This discrepancy results in position-dependent flux density scaling effects in Stokes I \citep{Hurley-Walker:2014v31p45,Hurley-Walker:2017A} and polarisation leakage in Stokes Q, U and V \citep{Sutinjo:2015v50p52S}. The most significant source of leakage is from Stokes I as this is where the sky signal dominates. In general, Stokes Q exhibits the strongest level of leakage but Stokes U and V can be contaminated with as much as $\sim$5\% leakage \citep{Lenc:2017} from Stokes I. For circular polarisation, such levels of leakage can result in false detections unless they are corrected for.

For a given epoch and frequency, the leakage pattern for a drift-scan beam will remain fixed for each of the Stokes parameters. The nature of the leakage pattern is a function of the calibrator location within the calibrator beam pointing, the beam pointing used for the calibrator scan and the beam pointing used for the drift scan. To overcome the limitations and errors associated with the analytic beam model we measured the effect of the beam on known GLEAM sources as they drift through the beam. It is important to note that deconvolution is not performed on the snapshot images. This ensured that the PSF (point-spread function) characteristics of the leaked component remains consistent between each of the Stokes parameters for a given snapshot. A secondary benefit is that the processing was greatly simplified; enabling real-time processing. As weaker GLEAM sources are more likely to be dominated by sidelobe confusion, we excluded them from the sampling process. For this reason, we only consider GLEAM field sources that have a peak flux density in Stokes I greater than 3 Jy\,PSF$^{-1}$. For each snapshot image, we sample Stokes I, Q, U and V at the pixel location of each GLEAM field source. 

\begin{figure*}
\centering
\includegraphics[width=0.32\linewidth]{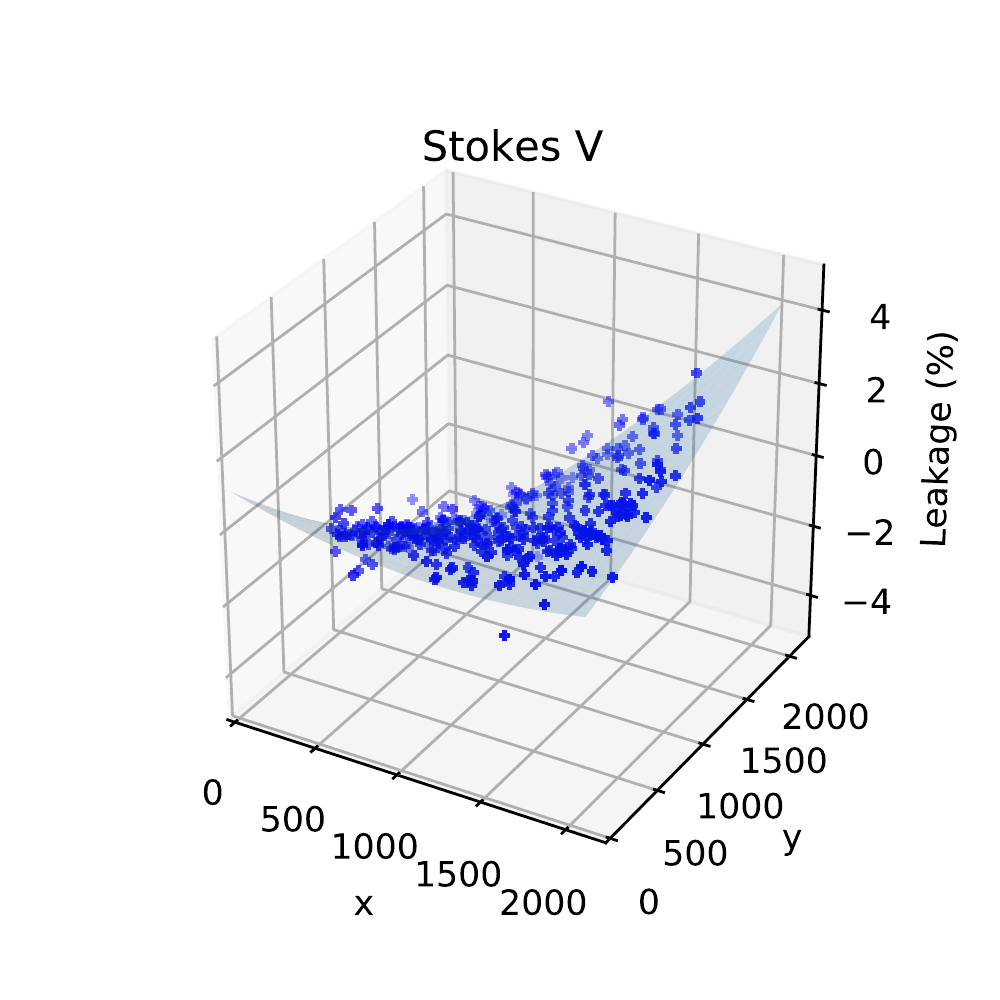}
\includegraphics[width=0.32\linewidth]{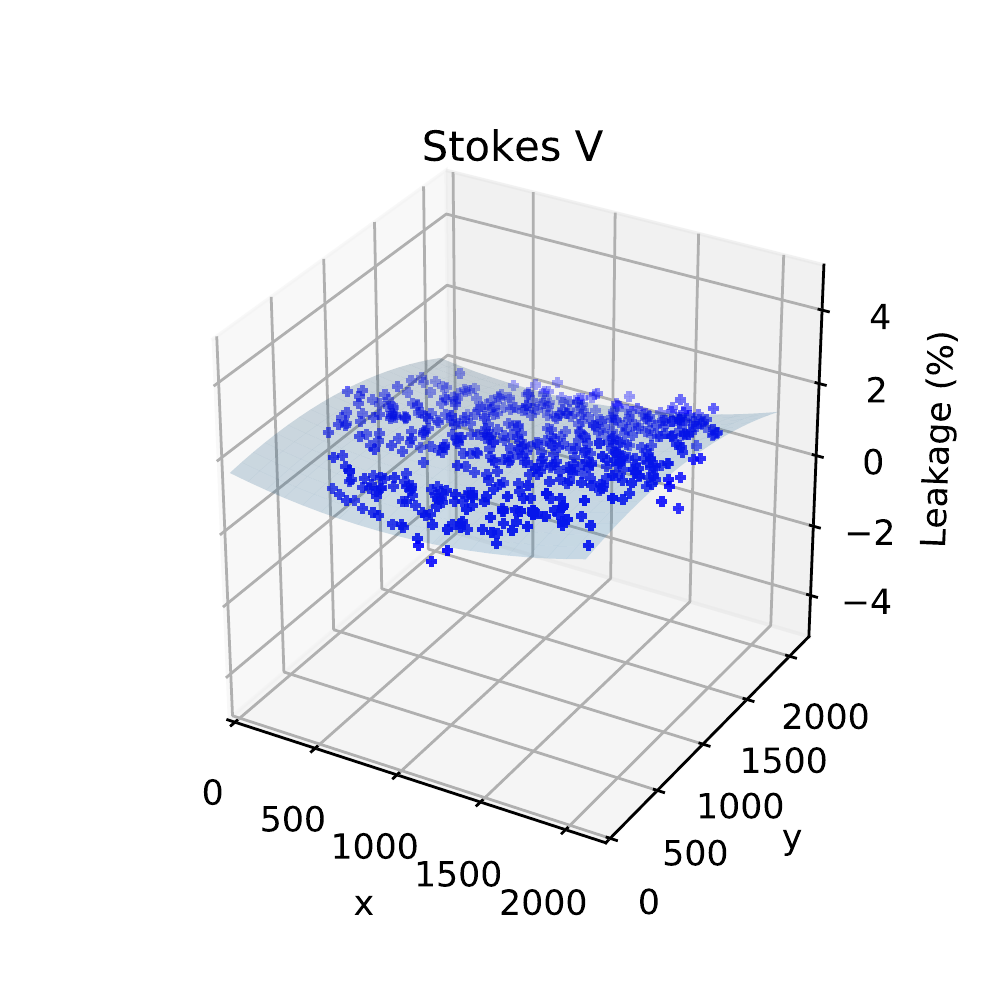}
\includegraphics[width=0.32\linewidth]{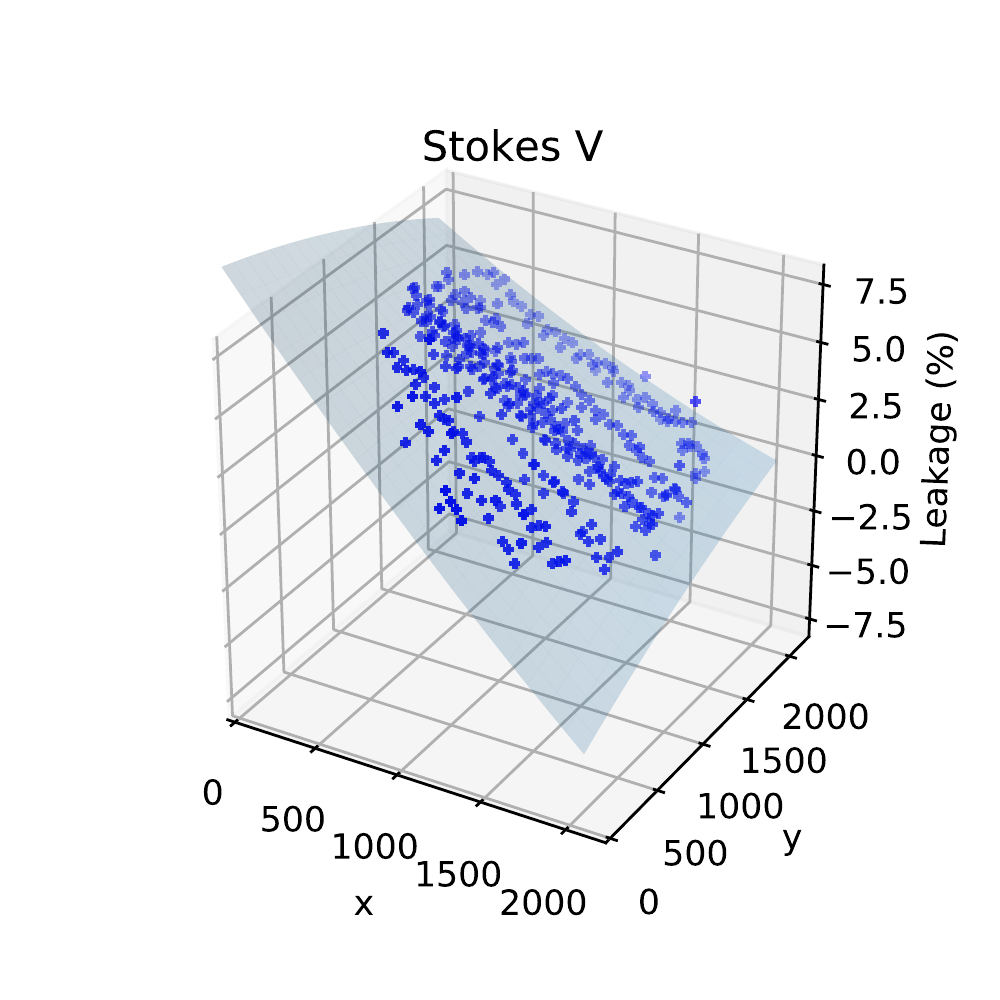}
\caption{Measured and fitted leakage in Stokes V in the 216 MHz band. The $x$ and $y$ axis are plotted in $(l,m)$ using units of pixels for a $25\arcdeg\times25\arcdeg$ field ($2\,187\times2\,187$ pixels), where $x$ represents the $l$ direction and $y$ represents the $m$ direction. Left: Observations taken from $+1.6\arcdeg$ drift scan on 2013-08-08. Centre: Observations taken from $-26.7\arcdeg$ drift scan on 2013-08-10. Right: Observations taken from $-72\arcdeg$ drift scan on 2013-08-18.}
\label{fig:leak215MHz}
\end{figure*}

To map the position-dependent leakage for the beam we assumed that all of the field sources are unpolarised. At low frequencies, this is a reasonable assumption \citep{Lenc:2016} and one that we verify later through our detection statistics. So any measured polarisation is assumed to result from beam errors. For circular polarisation we grid the measured fractional circular polarisation ($V/I$) at each of the sampled pixel locations. For small fields of view, a simple two-dimensional plane is sufficient to model the leakage across the beam, e.g. \citet{Lynch:2017b}, however for larger fields there is significant warping in the leakage behaviour and the fit errors increase significantly at the beam edges. To better model the leakage we fit a two-dimensional quadratic surface (over both spatial directions) to the grid in order to interpolate over the entire beam. Examples of this fitting are shown in Figure \ref{fig:leak215MHz} for drift scans at three different points on the meridian.

The leakage into any given Stokes parameter will be a mix of leakage from each of the other Stokes parameters. As the largest astrophysical signal is in Stokes I, it will dominate the observed leakage into each of the remaining Stokes parameters. So, for each snapshot, the previously fit leakage surface for Stokes V is used as position dependent scaling factor for the Stokes I map, the scaled Stokes I map is then subtracted from the Stokes V map to remove beam-associated leakage. The same process can also be repeated for Stokes Q and Stokes U but is not described in this paper. 

An aspect that is not taken into consideration is solving for XY-phase. Uncorrected XY-phase can result in leakage from Stokes U into Stokes V. Leakage of this form can lead to a false detection in Stokes V for a sufficiently strong linearly polarised source. To solve for XY-phase, at least one strong linearly polarised source would be required in each drift scan. At the time of this survey, such information was unavailable at long wavelengths. However, a survey of linearly polarised sources is currently being performed with the techniques developed here that will enable such calibration in future (Riseley in prep.). Based on prior observations at 154\,MHz we estimate the leakage from Stokes U to Stokes V to be of order $20-30\%$ \citep{Lenc:2016}. However, as linearly polarised sources detected with the MWA are generally weak at long wavelengths, typically $<5\%$ linearly polarised \citep{Bernardi:2013v771p105,Lenc:2016,Lenc:2017,OSullivan:2018}, $30\%$ leakage would typically result in less than $1.5\%$ of excess signal in Stokes V. Furthermore, for our Stokes V continuum observations, this would only be a potential source of error for sources with particularly low rotation measures ($\lvert\text{RM}\rvert<3$ rad\,m$^{-2}$) as they would otherwise be bandwidth depolarised by $>75\%$ over the 61.44\,MHz available bandwidth.

\subsection{Flux calibration}

Errors between the analytic beam model and the true beam can result in position-dependent flux calibration errors. During GLEAM survey processing, this was noted as a declination-dependent effect \citep{Hurley-Walker:2014v31p45,Hurley-Walker:2017A}, mainly because the mosaicking process dampened the effect in Right Ascension. However, a model of the flux calibration error can be formed using a similar process as the that used to model the leakage.

To model and correct for the position-dependent flux calibration errors, rather than measuring leakage, the scaling difference between the known GLEAM flux density for a source and the measured position-dependent flux density is gridded to form a scaling map. This scaling map is then applied to both Stokes I and Stokes V images of each snapshot to correct the flux density of field sources.

\subsection{Mosaic creation}

Mosaic creation was performed using the software package \textsc{swarp} \citep{Bertin:2002}. A three-stage process was utilised to generate the all-sky mosaics. In the first stage, individual mosaics were formed for each epoch and each observing band. This allowed the quality of the drift scans to be assessed and also avoided limitations of the software associated with the number of individual images that could be mosaicked simultaneously. The second stage combined the individual mosaics for a given frequency band into an all-sky mosaic. Finally, in the third stage, the two bands were combined to form the deep all-sky mosaic.

For the first stage of mosaicking, beam maps were created for each snapshot. During mosaicking, the individual snapshots were weighted against the square of the beam maps to minimize edge effects associated with noise spikes and increased error in the leakage corrections. Mosaics were formed using \textsc{swarp} with the corrected Stokes I and Stokes V snapshots. Mosaics were also formed using the uncorrected Stokes I and Stokes V snapshots to ultimately allow assessment of the effectiveness of the corrections in the final mosaics. All mosaics were formed with zenith-equal-angle projection and combined weight files were created by \textsc{swarp} for each of the drift scan mosaics and for each observing band.

For the second stage of mosaicking we used \textsc{swarp} to combine the individual drift-scan mosaics in each observing band. The combined weight files generated by \textsc{swarp} in the first stage were used as image weights during the mosaicking process. The final mosaics for each band were formed with zenithal equal area (ZEA) projection.

\begin{figure*}
\centering
\includegraphics[width=\linewidth]{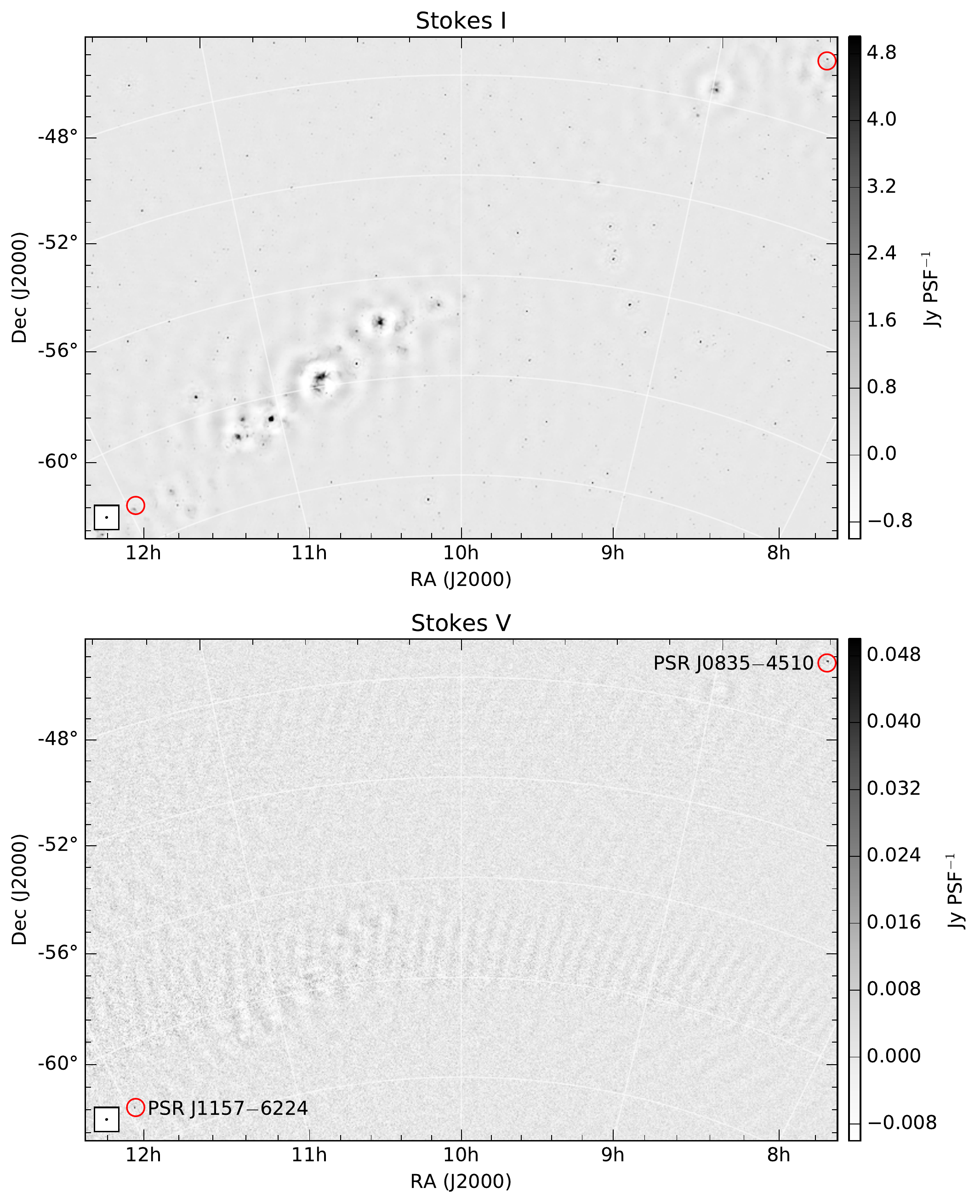}
\caption{A representative sub-set of the all-sky survey showing a Galactic region in Stokes I and Stokes V at 200 MHz. Two circularly polarised pulsars (circled in red) are detected in this region: PSRs J0835$-$4510 and J1157$-$6224. The approximate synthesised beam is shown inset and is $\sim$$3\arcmin$ in extent. SIN projection was used to generate this map.}
\label{fig:psr2}
\end{figure*}

In the final stage of mosaicking, the $169-200$\,MHz and $200-231$\,MHz mosaics were averaged to form a $200$\,MHz deep mosaic. This resulted in all-sky maps for the uncorrected Stokes I and Stokes V, and for the corrected Stokes I and Stokes V. Figure \ref{fig:psr2} shows a cut-out from the all-sky map showing a Galactic region in the corrected Stokes I and Stokes V maps. Two pulsars, PSRs J0835$-$4510 and J1157$-$6224, are clearly detected in the Stokes V map.

\subsubsection{Noise characterisation}
\label{sec:noise}

The final Stokes V image mosaics contain position-dependent variations in image noise. The main contributing factors to the levels of regional noise are the number of overlapping snapshots in that region, the effectiveness of calibration for the different epochs that contribute to that region (which is a function of the brightness and elevation of the calibrator), and the effectiveness of the leakage subtraction in that region.

At extreme declinations, i.e. $+18.6\arcdeg$ and $-72\arcdeg$, there are no overlapping drift scans to help improve sensitivity and so these field edges have higher noise levels. Some hour angles have higher levels of overlap between drift scans at the same declination and this leads to improved sensitivity in these regions e.g. between the 0$-$8\,h scans and the 6$-$16\,h scans there are only 2\,h of overlap, whereas between 6$-$16\,h scans and the 12$-$22\,h scans there are 4\,h of overlap.

Regions around bright Stokes I sources can also contribute towards increased Stokes V noise in regions where the leakage modeling is not as effective and in regions where there are extremely bright sources. Even if leakage is reduced to an ambitious level of $0.1\%$, a 100 Jy source would contribute 100 mJy to Stokes V. Since dirty images are used in processing, PSF sidelobes from these sources would contribute to noise over an extended region around each bright source. Hence we expect increased levels of Stokes V noise around bright sources such as the Crab Nebula and Pictor A.

\begin{figure*}
\centering
\includegraphics[width=\linewidth]{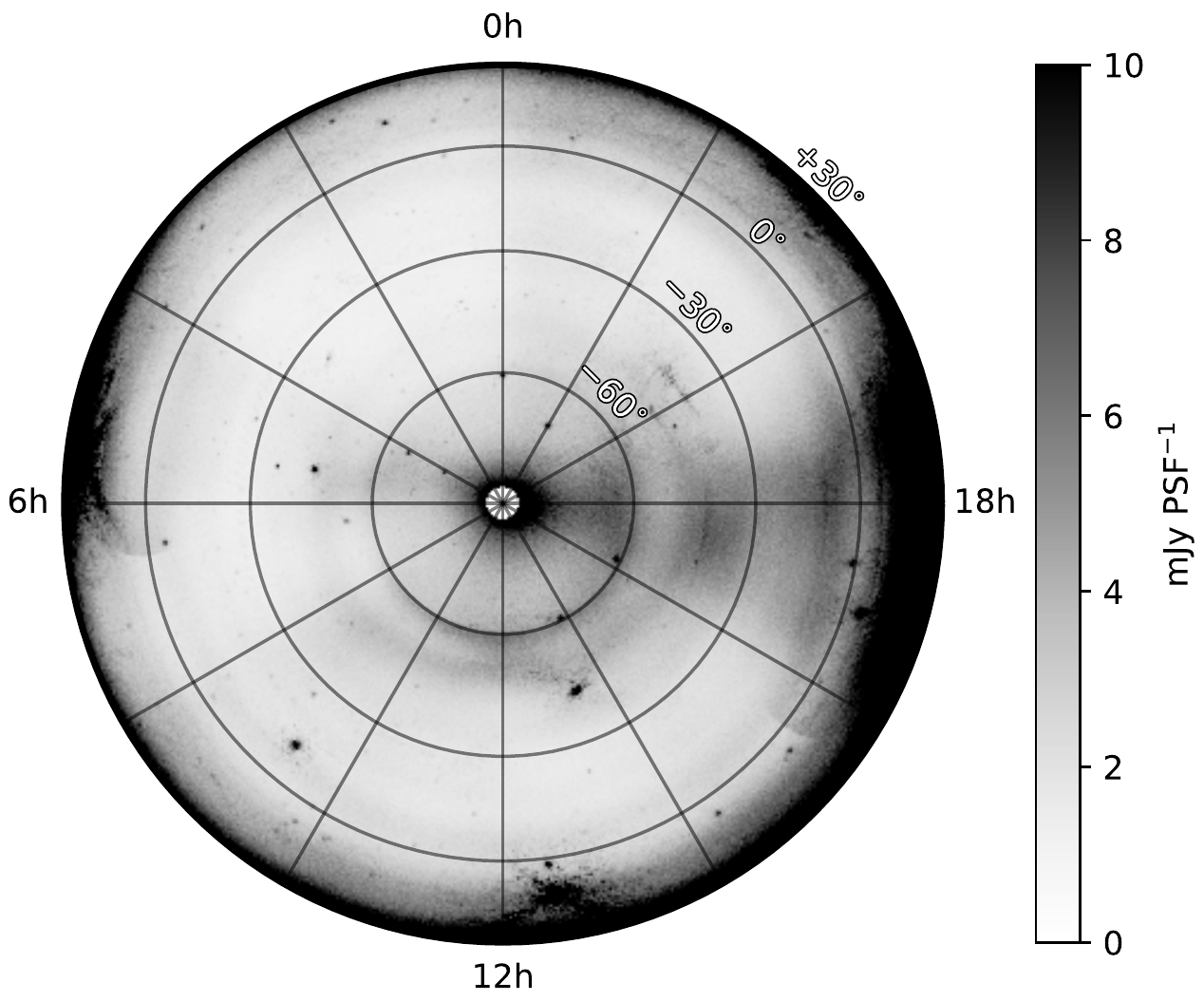}
\caption{All-sky map of measured RMS image noise in circular polarisation in the $200\,$MHz deep mosaic. A $20\times20$ pixel sliding window was used on the Stokes V mosaic to estimate image noise. Zenithal equal area (ZEA) projection is used, centred on a declination of $-90\arcdeg$.}
\label{fig:rms200MHz}
\end{figure*}

To map local noise, a $20\times20$ pixel sliding window was used. For each region within the mosaic, the mean and standard deviation is measured, any pixels with a mean subtracted peak exceeding 3$\sigma$ are excluded, the standard deviation and mean are then measured for the remaining pixels within the sliding window and recorded for that region. The resultant product is a local RMS map as shown in Figure \ref{fig:rms200MHz}. For the majority of the observable sky below $+10\arcdeg$ and above $-85\arcdeg$ the noise levels are typically of order 3\,mJy\,PSF$^{-1}$. There is a slight excess in noise at around 18\,h, this is primarily as a result of the bulk of the Galactic plane passing through this region, however, there is also limit hour-angle coverage in this region and this too will reduce sensitivity in this region.

\begin{figure}
\centering
\includegraphics[width=\linewidth]{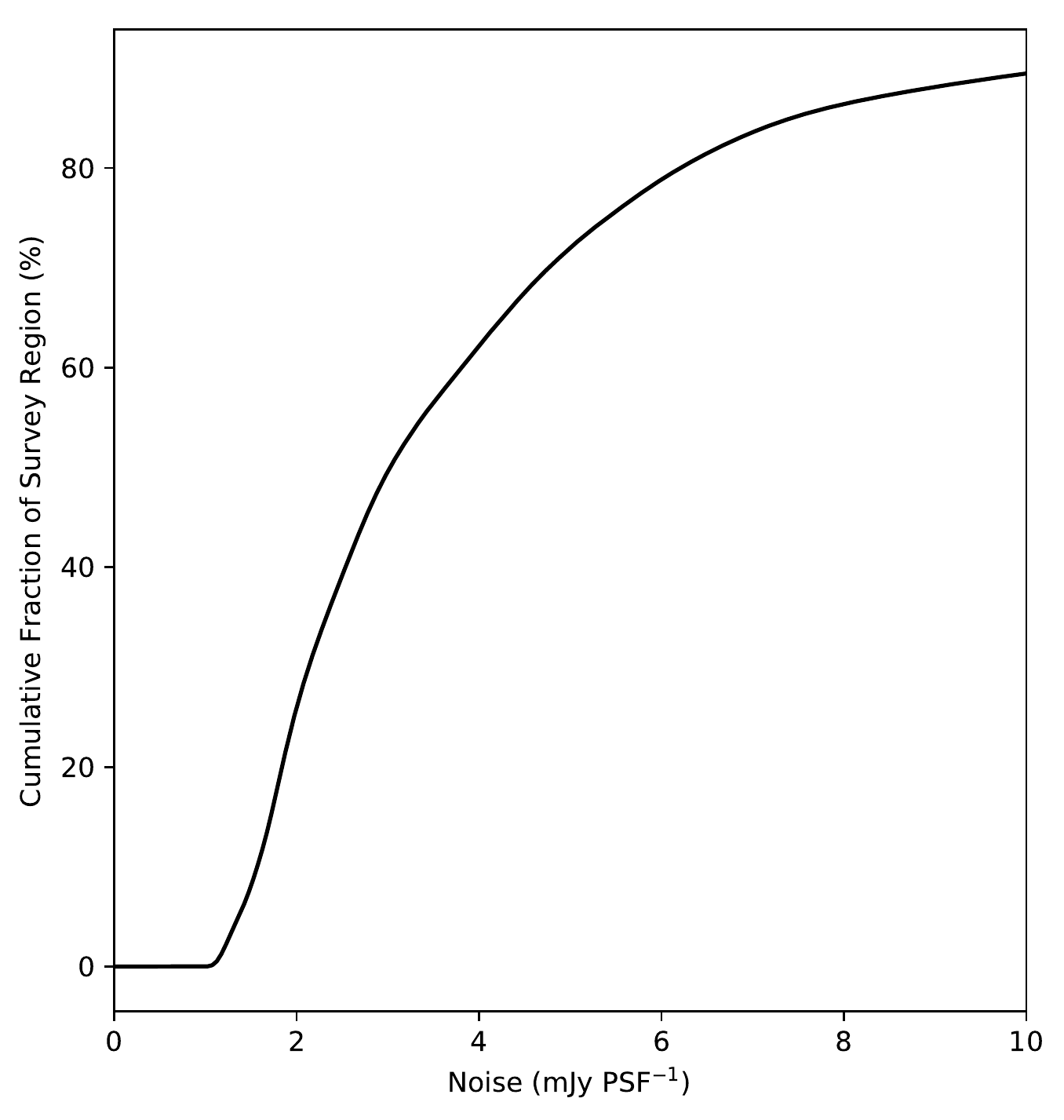}
\caption{A cumulative histogram showing the fraction of the survey region achieving a given Stokes V image noise limit at 200 MHz. The highest sensitivity achieved is $\sim$1\,mJy\,PSF$^{-1}$ and $\sim$$10\%$ of the survey region has noise exceeding $\sim$10\,mJy\,PSF$^{-1}$ (mostly constrained to the edge of the survey region, see Figure \ref{fig:noise200MHz}.).}
\label{fig:noise200MHz}
\end{figure}

Figure \ref{fig:noise200MHz} shows the proportion of the surveyed region that achieves a given noise level. The overall survey has a median noise level of 3.0\,mJy\,PSF$^{-1}$. Approximately $25\%$ of the surveyed region achieves a sensitivity better than 2\,mJy\,PSF$^{-1}$ and $75\%$ is better than 5.4\,mJy\,PSF$^{-1}$. This is a factor of $2-5$ improvement compared to the 10 mJy\,PSF$^{-1}$ sensitivity of GLEAM \citep{Hurley-Walker:2017A} in Stokes I with the same weighting scheme.

To verify the Gaussian nature of the noise statistics we considered all mosaic pixels which sit in regions where the noise was estimated to be within $3.0\pm0.5$\,mJy\,PSF$^{-1}$. These were fit with a Gaussian with a mean of $0.001$\,mJy\,PSF$^{-1}$ and a standard deviation of $3.029$\,mJy\,PSF$^{-1}$. The noise statistics are highly Gaussian with $95.39\%$ of pixels within $2\sigma$, $99.67\%$ within $3\sigma$, $99.986\%$ within $4\sigma$, and $99.9994\%$ within $5\sigma$ of the mean. Assuming Gaussian statistics, we would expect $\sim$$78$ false detections at the $4\sigma$ level, $\sim$$1$ at the $5\sigma$ level, and $\ll1$ at the $6\sigma$ level over the entire survey area.

\subsubsection{Flux density scale assessment}
\label{sec:fluxscale}

To assess the effectiveness of the flux-scale calibration, the flux density of all GLEAM sources with a cataloged peak brightness greater than 3 Jy\,PSF$^{-1}$ at 200\,MHz were measured in the uncorrected and corrected Stokes I mosaics. GLEAM sources in regions where the signal-to-noise was less than 20 were rejected to reduce measurements that are likely to be significantly affected by edge noise or source sidelobes. In total, 1779 GLEAM sources were available for use as suitable probes.

\begin{figure}
\centering
\includegraphics[width=\linewidth]{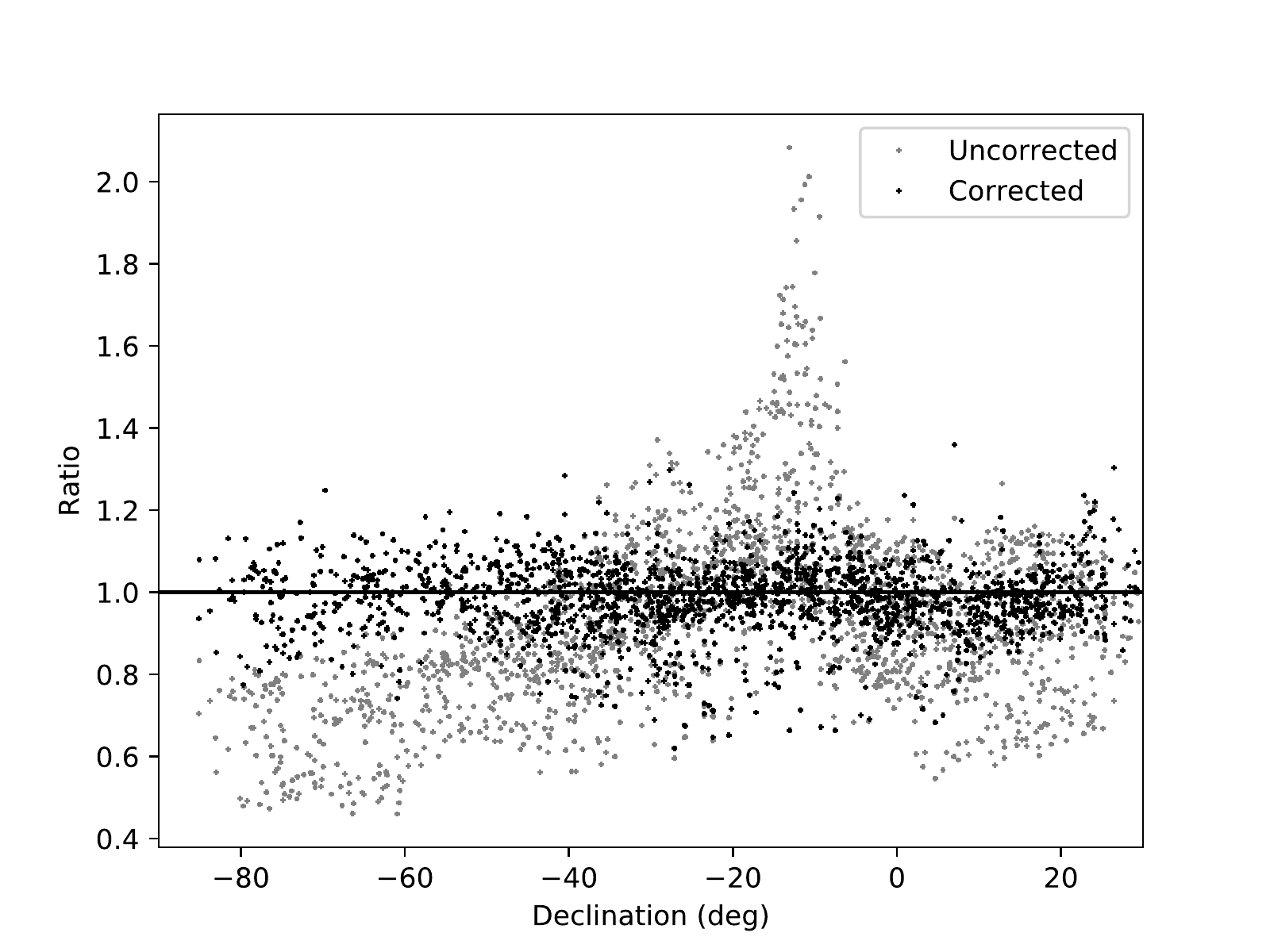}
\caption{Comparison of survey flux density scale of the 200 MHz deep mosaic with that of the GLEAM survey. Before correction there are declination-dependent effects. There are removed after correction.}
\label{fig:scale200MHz}
\end{figure}

Figure \ref{fig:scale200MHz} shows the ratio between the catalogued GLEAM source peak at 200 MHz and the measured Stokes I peak in the 200\,MHz mosaic plotted against declination for both the uncorrected and corrected maps. There is a clear declination dependence in the uncorrected data with an overall spread of $22\%$ in the measured flux densities and an absolute scaling of 0.943 compared to those of GLEAM. After correction, the declination dependence is no longer prominent with the overall spread reduced to $8.7\%$ and the absolute scaling at 0.984.

\begin{figure*}
\centering
\includegraphics[width=0.49\linewidth]{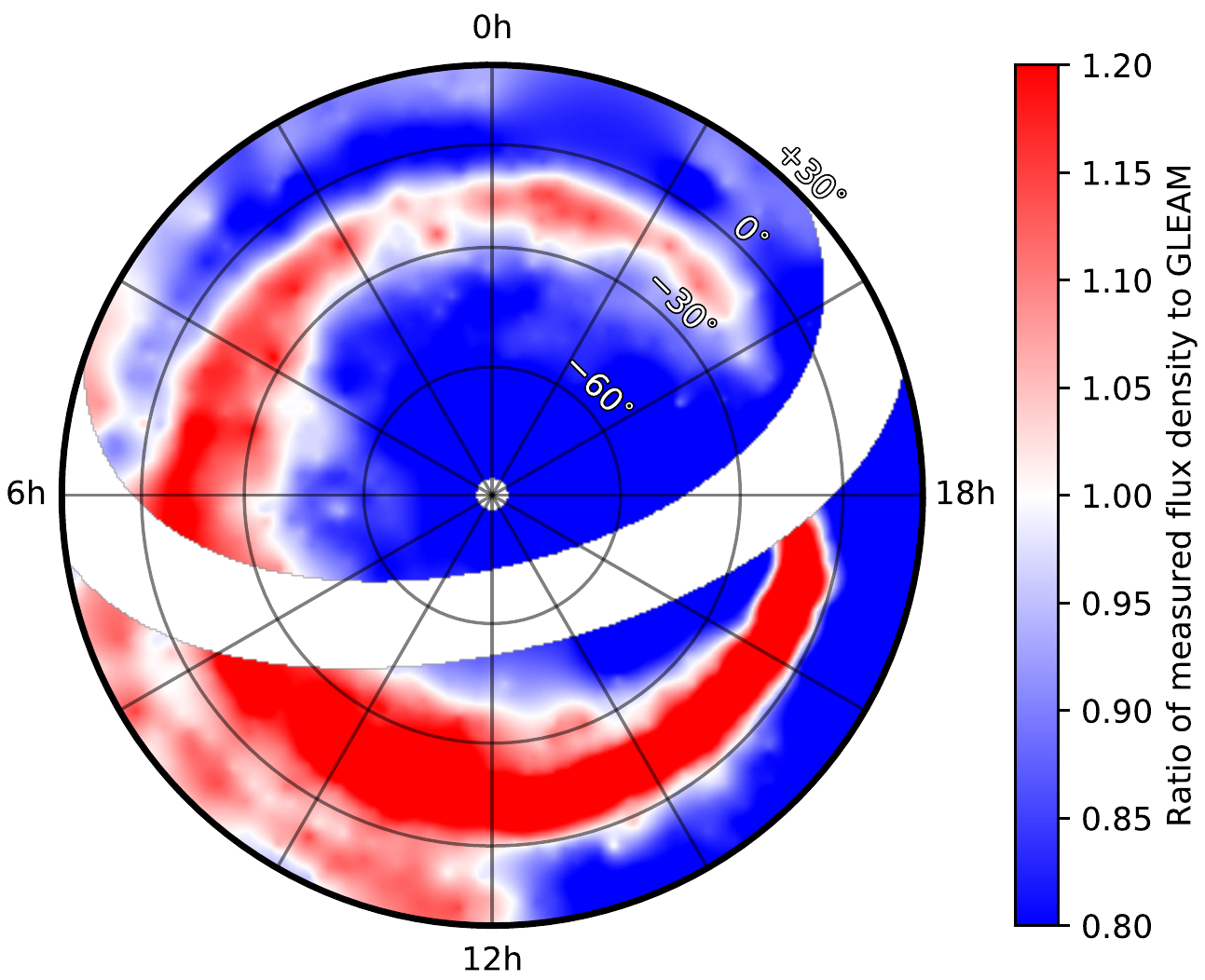}
\includegraphics[width=0.49\linewidth]{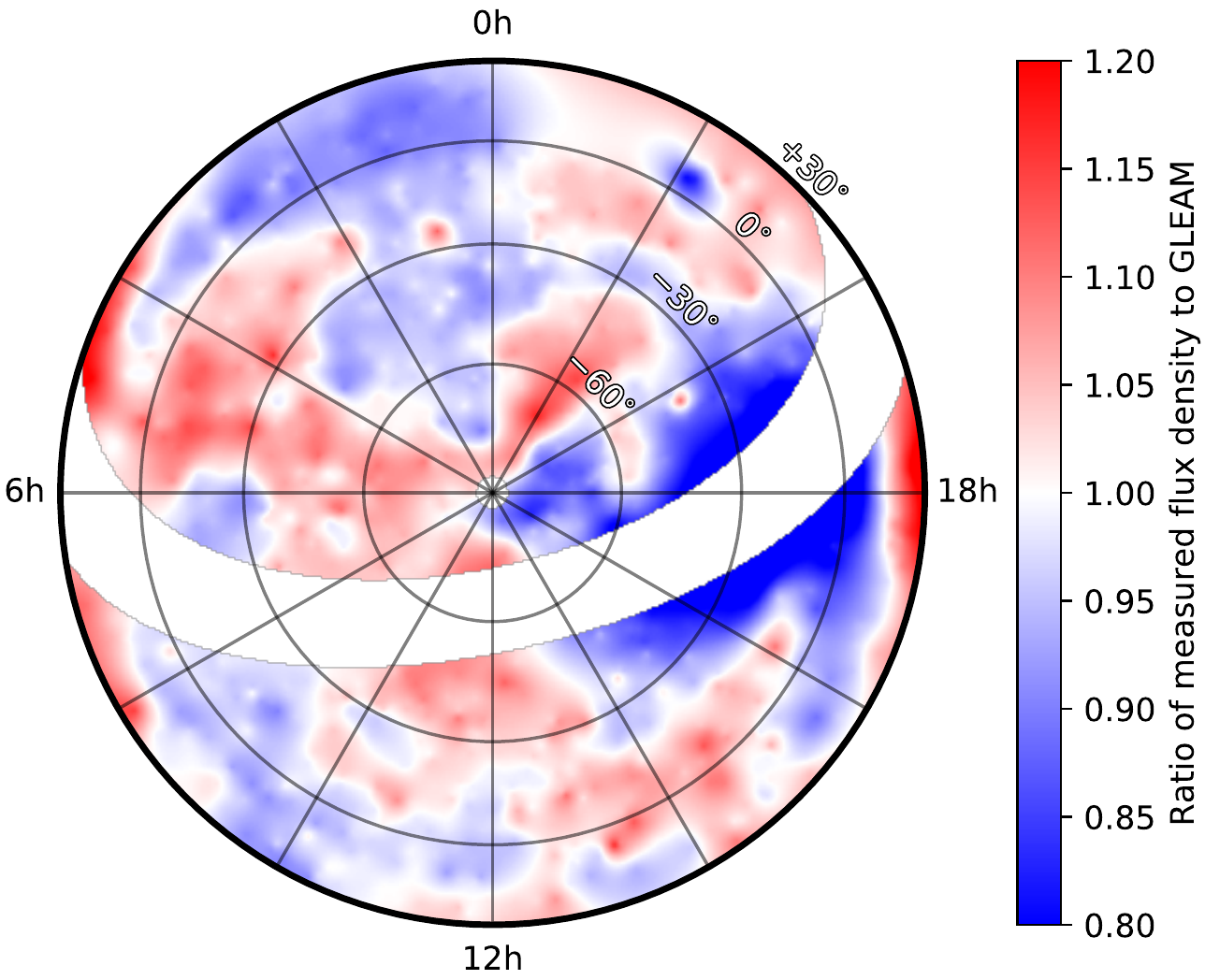}
\caption{Comparison of survey flux density scale of the 200 MHz deep mosaic with that of the GLEAM survey. The map on the left shows the flux density scale before correction and the map on the right shows the flux density scale after correction. In both maps the blanked strip removes the Galactic plane region as this was not surveyed by GLEAM. Zenithal equal area (ZEA) projection is used, centred on a declination of $-90\arcdeg$.}
\label{fig:flux200MHz}
\end{figure*}

Another representation of the flux density scale improvement is shown in Figure \ref{fig:flux200MHz}. This figure takes the same scaling measurements but maps the results as a function of sky position. The blanked region marks locations where no GLEAM measurements are available. The declination-dependent variations are clearly visible in the uncorrected maps, however, strong RA-dependent variations are also apparent e.g. an abrupt change from over-estimating to under-estimated the measured flux density at around 12h at high declinations. This RA-dependent effect results from calibrator source changes and also calibrator beam-former changes from epoch to epoch.

The greatest residual errors in the corrected flux density scale maps appear towards high declinations (particularly at the 18h and 5h mark) and towards the Galactic centre. The deviation at high declination is likely due to increased modelling errors at the edge of the map. At the edge of the map there are no further overlapping snapshots to help down-weight the increased fitting errors that are present there during mosaicking. The apparent underestimation of flux density towards the Galactic centre is likely due to sidelobe confusion affecting the sampling of source peak flux densities in that region. If it were a true underestimation of the flux density then it would affect the entire drift scan rather than just one part of it since the correction is used consistently over the entire drift scan.

\subsubsection{Leakage characterisation}
\label{sec:leakage}

Using a similar approach to that described in Section \ref{sec:fluxscale} for assessing the flux density scale, the 1779 GLEAM sources were also used to probe leakage at various sky locations. For each GLEAM source, the Stokes I and Stokes V flux density was measured at the location of that source.

\begin{figure}
\centering
\includegraphics[width=\linewidth]{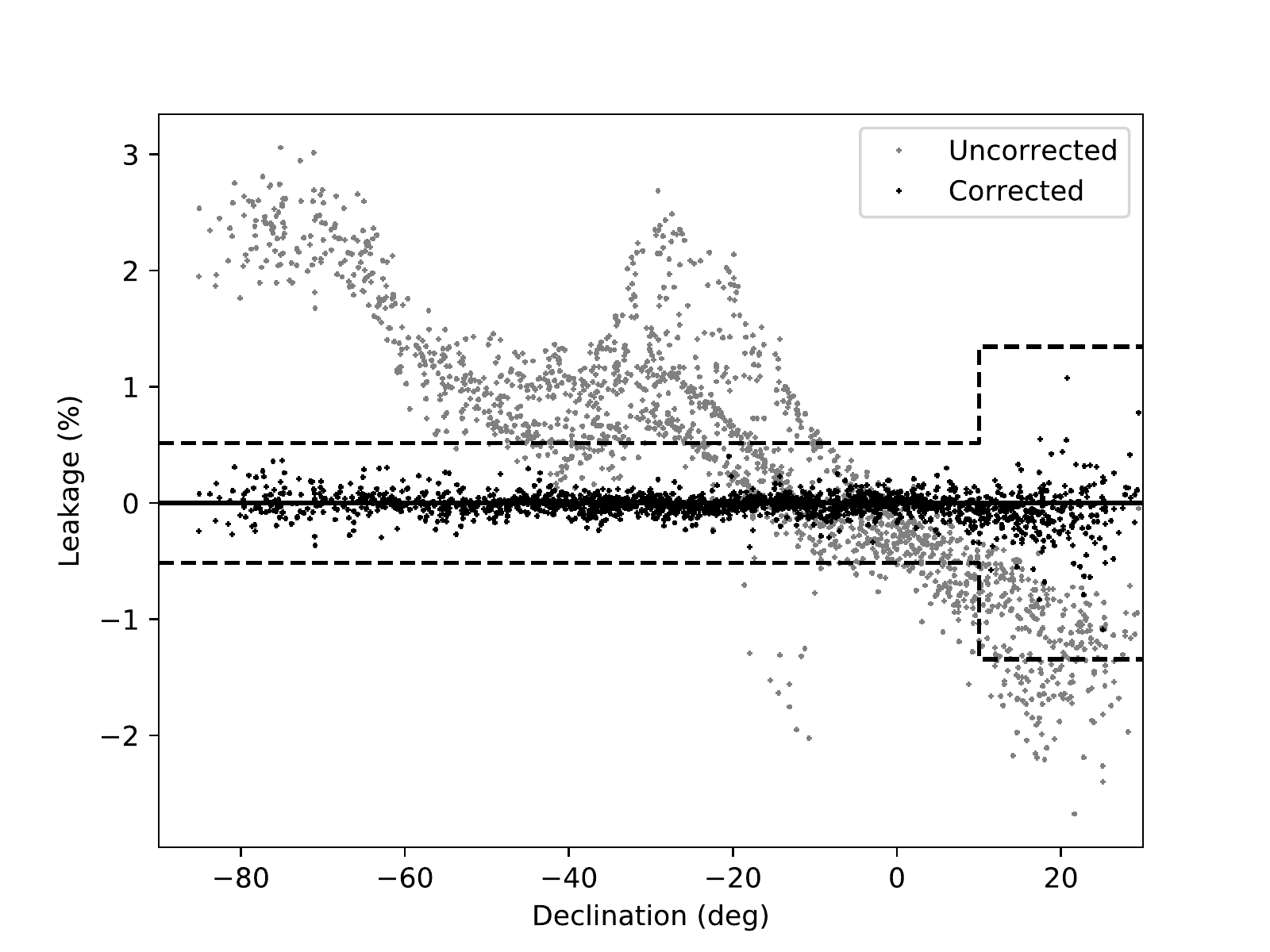}
\caption{Leakage from Stokes I to Stokes V plotted as a function of declination using GLEAM sources as points of reference in the 200 MHz deep mosaic. Before correction there are clear dependencies with declination whereas after correction the leakage is declination independent. Red dashed lines mark the 6$\sigma$ region of the scatter in Stokes V leakage after correction, where $1\sigma=0.12\%$ for declinations less than $+20\arcdeg$ and $0.3\%$ for declinations above $+20\arcdeg$.}
\label{fig:leak200MHz}
\end{figure}

Figure \ref{fig:leak200MHz} shows the percentage of Stokes I to Stokes V leakage as a function of declination before and after correction. Prior to correction the leakage showed a significant declination dependent behaviour with a typical spread of $\sim1\%$. At certain declinations there were several different bands of leakage. For example, between a declination of $-30\arcdeg$ and $-10\arcdeg$ three separate trends are apparent. These different trends, for the same declination, are caused by differences in epoch-to-epoch calibration.

After correction the declination dependence is removed and the typical spread is reduced to $0.12\%$. Since the drift scans at the highest and lowest declination are edge cases, i.e. they do not have overlapping drift-scans at higher and lower declinations, the model fit to the leakage pattern is not as well constrained. As a result, the spread of leakage increases in these regions. The effect is particularly pronounced above $+20\arcdeg$ declination where there are limited GLEAM sources to sample against (as a result of decreased sensitivity and regions not sampled by GLEAM). 

\begin{figure*}
\centering
\includegraphics[width=0.49\linewidth]{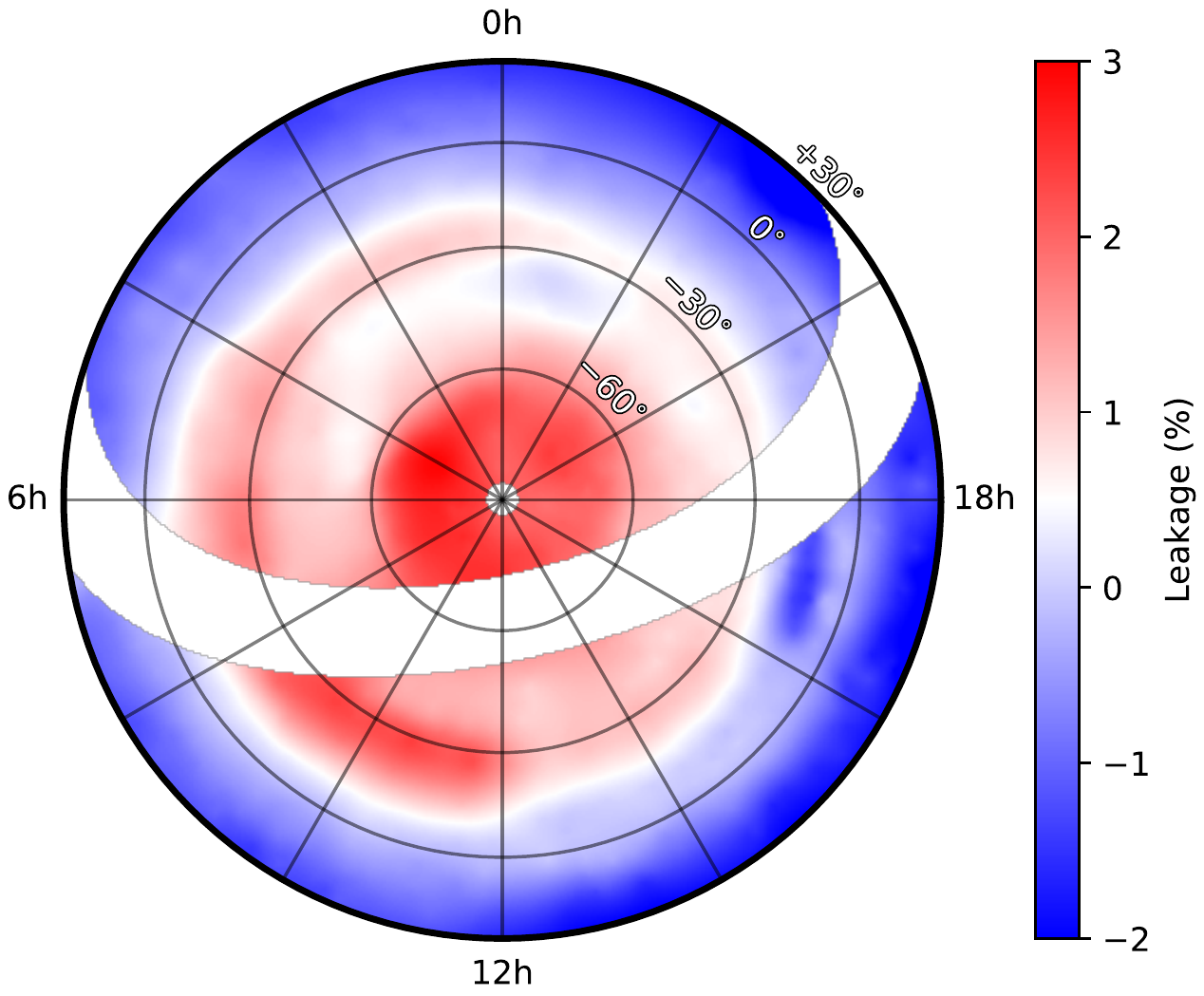}
\includegraphics[width=0.49\linewidth]{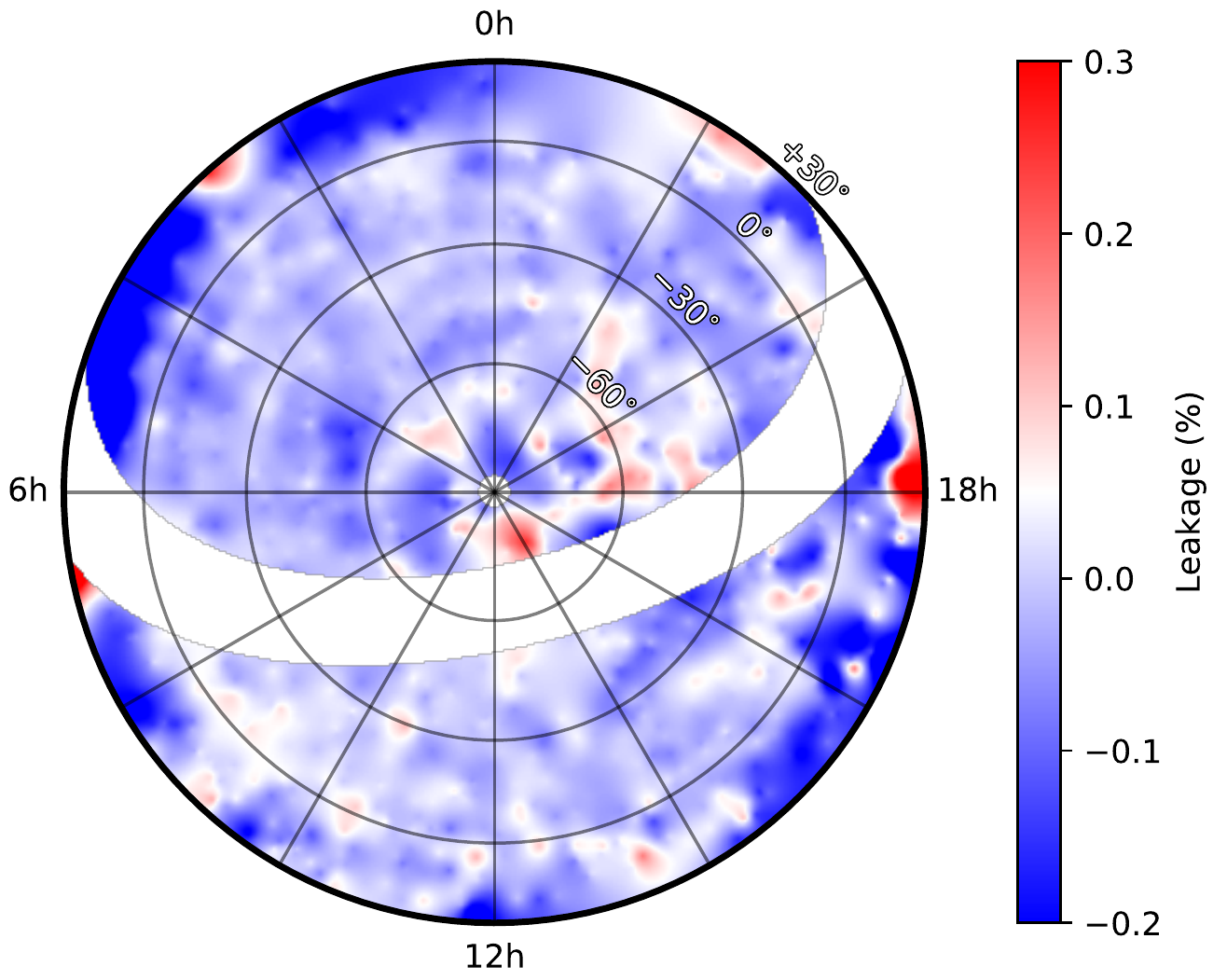}

\caption{Comparison of measured Stokes I to Stokes V leakage for GLEAM sources in the 200 MHz deep mosaic. The map on the left shows the leakage before correction and the map on the right after correction. In both maps the blanked strip removes the Galactic plane region as this was not surveyed by GLEAM. Note that the scale in the corrected map is an order of magnitude lower compared to the uncorrected map. Zenithal equal area (ZEA) projection is used, centred on a declination of $-90\arcdeg$.}
\label{fig:leakmap200MHz}
\end{figure*}

Mapping the leakage as a function of sky position, as shown in Figure \ref{fig:leakmap200MHz}, enables the characteristics of the leakage before and after correction to be analysed more readily. Before correction there are clear bands in declination where the leakage is either highly negative (high declinations) or highly positive (mid and low declinations). The epoch-to-epoch variations noted in Figure \ref{fig:leak200MHz}, which cause abrupt changes as a function of hour-angle, are seen more clearly, particularly at a declination of $\sim$$-30\arcdeg$.

In the corrected map, the overall leakage is improved by an order of magnitude and no clear trends are apparent for declinations below $\sim$$+20\arcdeg$. Above a declination of $\sim$$+20\arcdeg$, it is apparent that a slight excess of leakage is still present. As described in Section \ref{sec:fluxscale}, the model fitting is less constrained in this region and is affected by the reduced number of GLEAM sources that are available within this region. Nonetheless, the leakage is improved by an order of magnitude compared to the uncorrected map. 

\section{A Blind Survey}
\label{sec:results}

A blind search for circularly polarised sources was performed over the entire mosaicked map. The search recorded all pixels in the 200\,MHz Stokes V map that were six times greater in flux density than the associated RMS noise in that region based on the local RMS noise map (see Section \ref{sec:noise}). Islands of neighbouring detections were grouped as single detections. 

\begin{figure*}
\centering
\includegraphics[width=\linewidth]{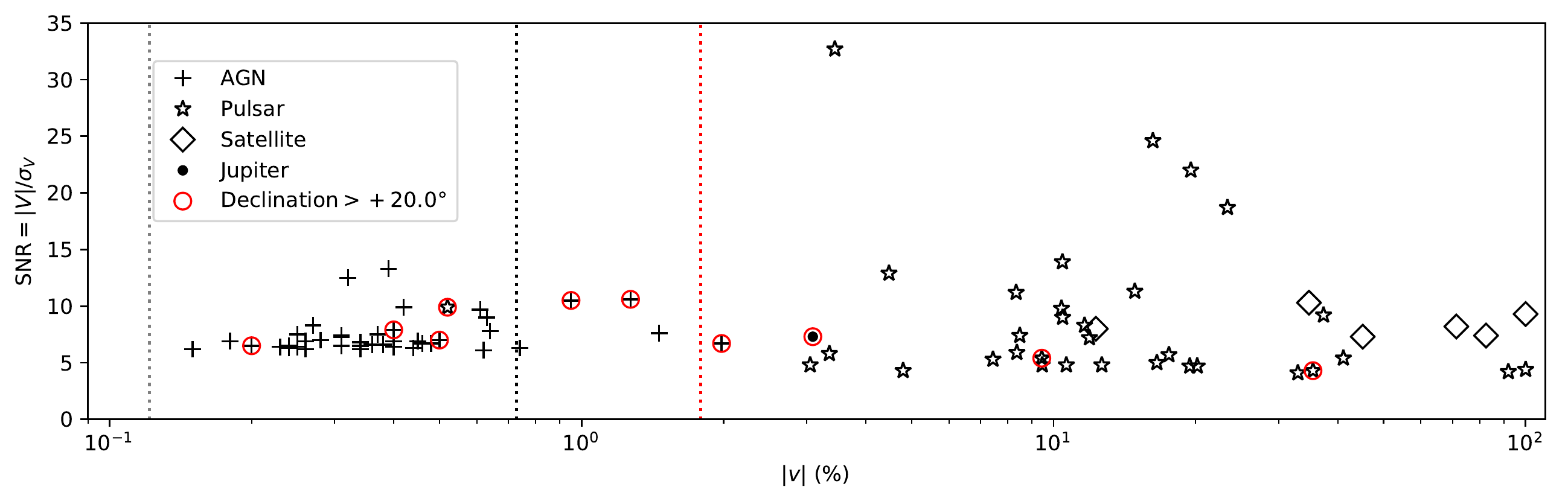}
\caption{Plot of all blind and targeted detections showing the signal-to-noise of the detection and the absolute percentage circular polarisation ($\lvert v\rvert$). The different source types are distinguished by marker symbol. Sources that are detected above a declination of $+20\arcdeg$ are circled. The grey dashed line shows the $1\sigma$ leakage in Stokes V for declinations below $+20\arcdeg$, the black dashed line shows the $6\sigma$ leakage in Stokes V for declinations below $+20\arcdeg$ and the red dashed line marks the $6\sigma$ leakage for declinations above $+20\arcdeg$.}
\label{fig:vvssnr}
\end{figure*}

In total, 63 unique detections were made above the $6\sigma$ level in the 200\,MHz Stokes V map. Of these, 41 are associated with bright radio galaxies, 15 are associated with known pulsars, one with Jupiter and six were associated with known artificial satellites. The significance of the detection, the fractional circular polarisation and the object type for all 63 sources are plotted in Figure \ref{fig:vvssnr}, the figure also includes targeted detections of pulsars which will be described in Section \ref{sec:pulsars}. To account for residual leakage that still exists after correcting for Stokes I to Stokes V leakage, a $6\sigma$ cut-off is applied to avoid false detections of "leaked" Stokes I sources. For declinations lower than $+20\arcdeg$ this cut-off is set to $0.72\%$ ($6\sigma$). For declinations greater than $+20\arcdeg$, where there is increased residual leakage, the cut-off is set to $1.8\%$ ($6\sigma$) to account for the increased residual leakage in that region. Applying this cut-off, one pulsar (associated with the extremely bright Crab Nebula with a flux density $>500$\,Jy) is rejected and all but three AGN. All remaining source detections are listed in Table \ref{tab:candidates}.

Table \ref{tab:candidates} also lists the angular separation between the blind detection and the nearest known radio source. For pulsars, the position is taken from the Australia Telescope National Facility (ATNF) Pulsar Catalog v1.56 \citep{Manchester:2005v129p1993}, AGN positions are taken from the NASA/IPAC Extragalactic Database (NED), and the position of Jupiter is determined for the 2013-11-11 epoch using the \textsc{pyephem} package\footnote{\url{http://rhodesmill.org/pyephem}}. Typical astrometric errors of $\sim$$0.3\arcmin$ are expected based on the PSF size and a signal-to-noise of $\sim$$6$. Uncorrected ionospheric effects will have a more dominant effect on astrometric error, contributing an additional error of order $\sim$$1\arcmin$ at $200\,$MHz \citep{Loi:2015v42p3707}.

\begin{table*}
\centering
\begin{tabular}{l r r r r r r r r r }
\hline \hline
Source   & RA              & Dec           & Ang. Sep.       & $V_{200}$ & SNR & $S_{200}$ (ref) & $v_{200}$ & $v_{\nu} $ & $\nu$ (ref) \\ 
         & (J2000)         &  (J2000)      & (arcmin) & (mJy)     &     & (mJy)       &     (\%)   &   (\%)    & (MHz)  \\ 
\hline
PSR J0034$-$0721  & 00$\rah$34$\ram$10$\ras$ & -07$\arcdeg$21$\arcmin$38$\arcsec$ & 0.3 &   +30.5 & 13.9 &   292.0 (M) & $+10.4$ & $+12.0$ &  149 (N)  \\
PSR J0437$-$4715  & 04$\rah$37$\ram$20$\ras$ & -47$\arcdeg$14$\arcmin$24$\arcsec$ & 1.0 &  +135.4 & 24.6 &   834.0 (M) & $+16.2$ &  $+8.0$ &  438 (Y)  \\
PSR J0630$-$2834  & 06$\rah$30$\ram$51$\ras$ & -28$\arcdeg$34$\arcmin$28$\arcsec$ & 0.5 &   -20.7 & 12.9 &   463.0 (M) &  $-4.5$ &  $-2.8$ & 1400 (J)  \\
PSR J0738$-$4042  & 07$\rah$38$\ram$31$\ras$ & -40$\arcdeg$41$\arcmin$56$\arcsec$ & 0.5 &   +14.0 &  7.4 &   165.0 (M) &  $+8.5$ &  $-3.7$ & 1400 (J)  \\
PSR J0742$-$2822  & 07$\rah$42$\ram$50$\ras$ & -28$\arcdeg$22$\arcmin$29$\arcsec$ & 0.1 &   -15.3 &  9.0 &   146.0 (X) & $-10.5$ &  $-2.4$ & 1400 (J)  \\
PSR J0745$-$5353  & 07$\rah$45$\ram$04$\ras$ & -53$\arcdeg$52$\arcmin$37$\arcsec$ & 1.5 &   -18.7 &  9.2 &             &         &  $+4.7$ & 1400 (J)  \\ 
PSR J0835$-$4510  & 08$\rah$35$\ram$22$\ras$ & -45$\arcdeg$10$\arcmin$20$\arcsec$ & 0.6 &  +243.6 & 32.7 &  7075.0 (M) &  $+3.4$ &  $-6.2$ & 1400 (J)  \\
PSR J1136$+$1551  & 11$\rah$36$\ram$04$\ras$ & +15$\arcdeg$50$\arcmin$49$\arcsec$ & 0.4 &  -159.7 & 18.7 &   684.0 (M) & $-23.4$ & $-17.0$ &  149 (N)  \\
PSR J1157$-$6224  & 11$\rah$57$\ram$26$\ras$ & -62$\arcdeg$24$\arcmin$06$\arcsec$ & 1.1 &   +50.8 & 11.3 &     342 (L) & $+14.9$ & $+13.2$ & 1400 (J)  \\
PSR J1327$-$6222  & 13$\rah$27$\ram$28$\ras$ & -62$\arcdeg$22$\arcmin$00$\arcsec$ & 1.4 &   -33.8 &  7.2 &   284.0 (M) & $-11.9$ &  $+7.1$ & 1400 (J) \\ 
PSR J1453$-$6413  & 14$\rah$53$\ram$40$\ras$ & -64$\arcdeg$12$\arcmin$31$\arcsec$ & 1.0 &   +57.0 & 11.2 &   684.0 (M) &  $+8.3$ &  $+6.9$ & 1400 (J)  \\
PSR J1651$-$4246  & 16$\rah$51$\ram$53$\ras$ & -42$\arcdeg$45$\arcmin$56$\arcsec$ & 0.6 &  -213.9 & 22.0 &  1095.0 (M) & $-19.5$ &  $-5.6$ & 1400 (J)  \\
PSR J1932$+$1059  & 19$\rah$32$\ram$15$\ras$ & +10$\arcdeg$58$\arcmin$47$\arcsec$ & 0.6 &   -58.3 &  8.3 &   501.0 (M) & $-11.6$ & $-22.8$ &  149 (N)  \\
PSR J2048$-$1616  & 20$\rah$48$\ram$35$\ras$ & -16$\arcdeg$16$\arcmin$30$\arcsec$ & 0.2 &   +17.6 &  9.8 &   169.0 (M) & $+10.4$ &  $+7.1$ & 1400 (J)  \\
\hline
PKS J0006$-$4235\textsuperscript{a} & 00$\rah$05$\ram$59$\ras$ & -42$\arcdeg$32$\arcmin$19$\arcsec$ & 2.4 &  $-12.8$ &  6.3 &  1718.0 (L) & $-0.74$ &   &   \\ 
PMN J0257$-$2433\textsuperscript{a} & 02$\rah$57$\ram$23$\ras$ & -24$\arcdeg$31$\arcmin$37$\arcsec$ & 2.4 &  $+10.1$ &  7.6 &   692.4 (L) &  $+1.5$ &   &   \\
3C 139.2\textsuperscript{a}         & 05$\rah$24$\ram$35$\ras$ & +28$\arcdeg$13$\arcmin$27$\arcsec$ & 1.9 & $-227.3$ &  6.7 &  11486.3 (L) & $-2.0$ &   &   \\
\hline
Jupiter\textsuperscript{a}          & 07$\rah$27$\ram$44$\ras$ & +21$\arcdeg$54$\arcmin$17$\arcsec$ & 0.4  & $-37.0$ &  7.3 &   1198.7 (L) & $-3.0$ &   &   \\  [1ex]
\hline
\multicolumn{10}{l}{\textsuperscript{a}\footnotesize{These sources may be affected by excessive leakage from Stokes I or Stokes U.}} \\
\end{tabular}
\caption{List of all sources detected above 6$\sigma$ at 200\,MHz in circular polarisation that have an associated astrophysical counterpart. The RA, Dec are the J2000 position of the peak in MWA 200 MHz mosaic image. The angular separation is the angular distance between source peak and the catalogued position of the nearest identified radio source (see text for details). $V_{200}$ is the measured Stokes V flux density. SNR is the signal to noise of the detected source. $S_{200}$ is the estimated total intensity at 200\,MHz. $v_{200}$ is the estimated fractional circular polarisation. $v_{\nu}$, and $\nu$ are the fractional circular polarisation and frequency found in literature. References provided within parenthesis refer to J:\citet{Johnston:2017}, L:This work, M:\citet{Murphy:2017}, N:\citet{Noutsos:2015}, X:\citet{Xue:2017}, and Y:\citet{You:2006}.}
\label{tab:candidates}
\end{table*}

\subsection{Pulsars}
\label{sec:blindpulsars}

\begin{figure*}
\centering
\includegraphics[width=\linewidth]{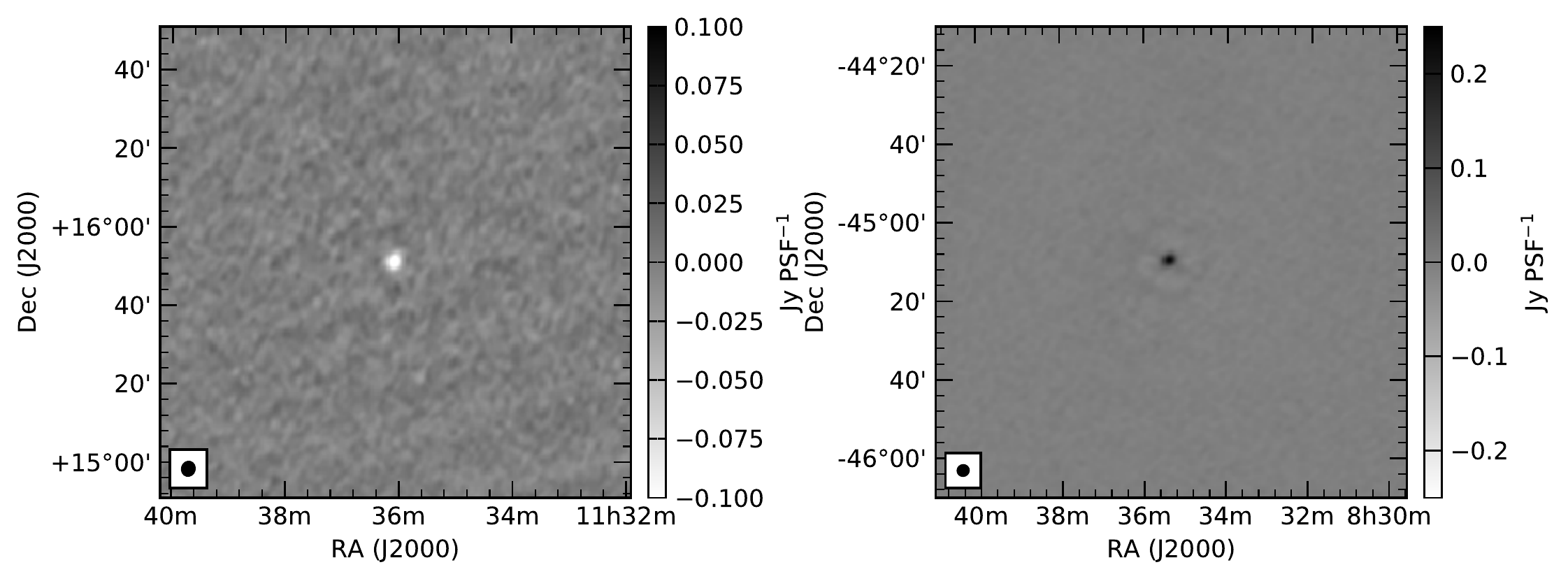}
\caption{Image of two sample pulsars from 200 MHz deep mosaic showing detections in different signs of circular polarisation. Left: PSR J1136$+$1551 (negative sign). Right: PSR J0835$-$4510 (positive sign). The synthesised beam is shown inset for PSRs J1136$+$1551 and J0835$-$4510, they are $3.5\times3.2$ (position angle $-15\arcdeg$) and $2.8\times2.7$ (position angle $-68\arcdeg$), respectively.}
\label{fig:pulsars200MHz}
\end{figure*}

In total, 14 pulsars were detected in the blind survey. Table \ref{tab:candidates} lists all of the detected pulsars, the observed characteristics at 200\,MHz and observations of circular polarisation from literature (where available). Images of two of the detected pulsars, PSRs J1136$+$1551 and J0835$-$4510, are shown in Figure \ref{fig:pulsars200MHz} and demonstrate that circular polarisation of either sign can be observed.

All of the detected pulsars have relatively high fractional circular polarisation at 200\,MHz ($>3\%$). In our sample, the sign of polarisation does not appear to be biased either way with near equal proportions having either negative or positive sign. Three pulsars (PSRs J0034$-$0721, J1136$+$1551, and J1932$+$1059) were previously observed with LOFAR at 149\,MHz \citep{Noutsos:2015} and exhibit a consistent sign of circular polarisation and similar fractional polarisation in our 200\,MHz observations. Four pulsars (PSRs J0738$-$4042, J0745$-$5353, J0835$-$4510, and J1327$-$6222) exhibit a sign flip at 200\,MHz compared to observations at 1.4\,GHz \citep{Johnston:2017}. PSR J0835$-$4510 is the most prominent of these given that it is detected with a signal-to-noise of greater than 30. PSR J0745$-$5353 is detected in circular polarisation at 200\,MHz but is not detected in Stokes I.

Of the 60 pulsars detected in the GLEAM 200\,MHz survey data \citep{Murphy:2017}, we detect 11 at the $6\sigma$ level - a proportion of $\sim$$18\%$. We also detect an additional three pulsars which were in regions not explored by GLEAM (e.g. the Galactic plane) or were too faint to be seen in the confusion limited Stokes I maps.

\subsection{AGN}
\label{sec:agn}
It is unusual to find AGN with a fractional circular polarisation greater than $0.5\%$ so the three remaining AGN were examined in more detail. All three AGN are detected just above our $6\sigma$ threshold for residual leakage and the $6\sigma$ local noise threshold. Subtle local variations in noise and/or systematic leakage may have been sufficient to push these sources above the threshold. 3C 139.2 itself is situated at the edge of the surveyed region at a declination of $\sim$$+28\arcdeg$ where sensitivity is extremely poor. Being situated at both the edge of the field and near the Galactic plane (RA$=\sim$$5.4$\,h) also limits the effectiveness of leakage subtraction at the source location. Inspection of the 216\,MHz mosaic confirms that leakage subtraction was particularly poor at that location and so is likely a false detection.

The two remaining AGN (PMN J0257$-$2433 and PKS J0006$-$4235) have peaks in circular polarisation that are significantly offset ($>2\arcmin$) from the Stokes I peak, offsets that are significantly higher than expected based on the SNR, PSF and ionosphere. While these may be associated with a chance alignment by a foreground circularly polarised source it is more likely that these are associated with AGN hot spots. If these hotspots are linearly polarised and exhibit a low rotation measure (RM) they may be symptomatic of leakage from Stokes U to Stokes V. Such leakage results from an uncorrected XY-phase and has been observed to occur with low-RM sources \citep{Lenc:2017} with a fractional leakage of $\sim$$20\%$. \citet{Taylor:2009v702p1230} determined PMN J0257$-$2433 was $5\%$ linearly polarised with an RM of $11.7\pm4.5$\,rad\,m$^{-2}$ at 1.4\,GHz. If the source exhibits similar characteristics at 200\,MHz then Stokes U to Stokes V leakage would be sufficient to cause a false detection in this instance. The same may be true for PKS J0006-4235 as it is morphologically similar. The source is known to be a 20 GHz source \citep{Murphy:2010} but its polarimetric characteristics are not known. Further observations of this source would be required to determine its true nature.

\subsection{Jupiter}
\label{sec:jupiter}
While Jupiter is known to exhibit a fractional circular polarisation $\sim$$1\%$ level at 3.24\,GHz \citep{Seaquist:1969}, we measure $\sim$$3.1\%$ ($7.3\sigma$) at 200\,MHz. Jupiter was at a relatively high declination ($+21.9\arcdeg$) and close to the Galactic plane (7.9\,h) in the epoch where it was detected. As with the AGN examined in Section \ref{sec:agn}, the source is in a region where sensitivity is poor and residual leakage is high and this may have resulted in an over-estimation or even a false detection. Further observations would be required to confirm this.

\subsection{Artificial satellites}

\begin{figure*}
\centering
\includegraphics[width=0.8\linewidth]{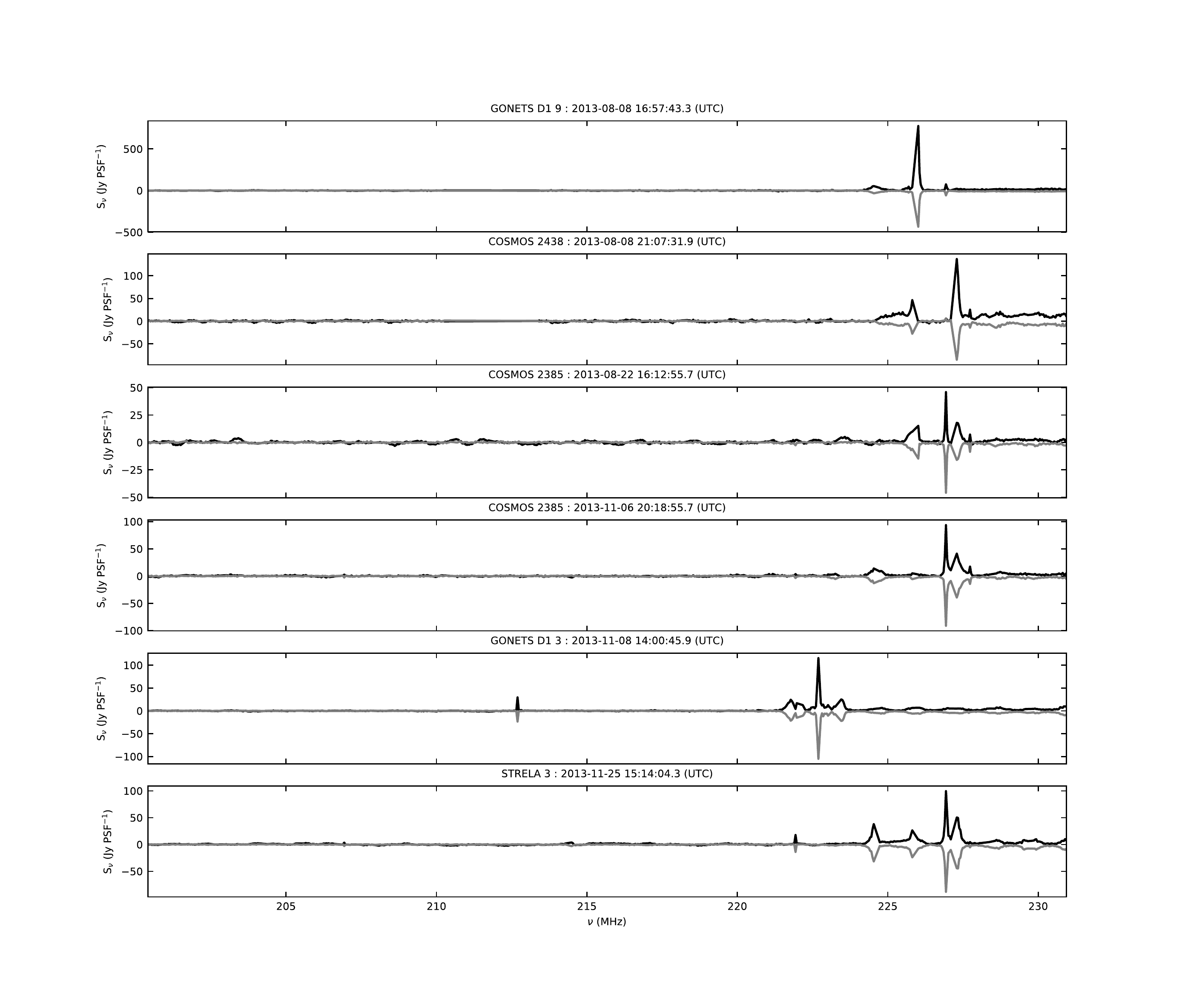}
\caption{The measured spectral energy distribution of all point-like ``transients'' observed in the survey data. Stokes I is shown in black and Stokes V in red. The unusual profile and high degree of circular polarisation suggests that these are most likely due to intermittent transmissions from artificial satellites that pass through the observed field.}
\label{fig:satrfi}
\end{figure*}

Six of the detections in circular polarisation at 200 MHz were not associated with any astrophysical sources. Upon closer inspection of the original snapshot images that were integrated to form the all-sky mosaic, it was discovered that each of the detected sources only appeared in a single 2-minute snapshot. Further investigation of the original spectral cube for each of the snapshots revealed that the spectral energy distribution for each of the transient sources exhibited narrow-band spikes, see Figure \ref{fig:satrfi}, that are commonly associated with radio frequency interference.

It is known that satellites can reflect FM-band signals from the Earth and this can be detected in MWA observations \citep{Tingay:2013b}. The result is typically a moving point source that tracks with the location of the satellite as it moves through its orbit. However, the detections found here were in a band that is well above the FM band. Secondly, the detections are unresolved and do not appear to move, this suggests that this is as a result of direct short-term transmission from the satellite itself rather than reflection.

To confirm the satellite nature of the detections, satellite ephemeris was obtained from \url{http://space-track.org/}. The positions of satellites were tracked over the 2-minute period of each snapshot to determine if any corresponded with the position of the detected source. In each instance a satellite was identified within $1\arcmin$ of the ``transient'' source, these are listed in Table \ref{tab:sat}. The fields were re-imaged at higher time-resolution around the estimated location of a nearby satellite to confirm the association of the flare with the satellite and the time of the flare was also recorded in Table \ref{tab:sat}.

\begin{table*}
\centering
\begin{tabular}{l l l l l l }
\hline \hline
Satellite & NORAD ID & RA & Dec & Ang. Sep. & Time \\ 
 &  & (J2000) & (J2000) & (arcmin) & (UTC) \\ 
\hline
GONETS D1 9 & 27060 & 21$\rah$59$\ram$06$\ras$ & +14$\arcdeg$02$\arcmin$30$\arcsec$ & 0.7 & 2013-08-08 16:57:43.3 \\
COSMOS 2438 & 32955 & 02$\rah$20$\ram$23$\ras$ & +10$\arcdeg$50$\arcmin$47$\arcsec$ & 0.3 & 2013-08-08 21:07:31.9 \\
COSMOS 2385 & 27056 & 21$\rah$52$\ram$28$\ras$ & -05$\arcdeg$17$\arcmin$59$\arcsec$ & 0.6 & 2013-08-22 16:12:55.7 \\
COSMOS 2385 & 27056 & 07$\rah$47$\ram$45$\ras$ & -42$\arcdeg$28$\arcmin$46$\arcsec$ & 0.5 & 2013-11-06 20:18:55.7 \\
GONETS D1 3 & 23789 & 00$\rah$44$\ram$36$\ras$ & -51$\arcdeg$00$\arcmin$56$\arcsec$ & 0.9 & 2013-11-08 14:00:45.9 \\
STRELA 3    & 37153 & 03$\rah$49$\ram$00$\ras$ & -31$\arcdeg$31$\arcmin$41$\arcsec$ & 0.7 & 2013-11-25 15:14:04.3 \\ [1ex]
\hline
\end{tabular}
\caption{Table of artificial satellite emission detected in circular polarisation. The coordinate and time of the satellite are listed at the time of emission. The angular separation is between the observed emission and the predicted location of the satellite based on the satellites ephemeris.}
\label{tab:sat}
\end{table*}

\section{Targeted Survey}
\subsection{Pulsars}
\label{sec:pulsars}

As noted in Section \ref{sec:blindpulsars}, the pulsars detected in the blind survey all have relatively high fractional circular polarisation. This makes them excellent candidates for a targeted survey. Of the 2\,613 known pulsars in ATNF Pulsar Catalog (v1.56), 2\,376 of these are within our survey region. A targeted survey of these pulsars was performed by probing the circular polarisation map at each of the pulsar locations (as recorded in the ATNF Pulsar Catalog) for significant emission above the esimated local image noise. For the targeted survey, a lower ($4\sigma$) threshold was utilised compared to the blind survey since an a priori position was known.

Table \ref{tab:pulsar} lists all pulsars that were detected above the $4\sigma$ threshold, the table excludes pulsars that were already detected in the blind survey. In total, 32 pulsars were detected, 18 of these were not detected by the blind survey. All of the pulsars with low frequency circular polarisation previously reported in literature have consistent sign with the detections reported here. Three of the pulsars have a sign flip in circular polarisation compared to measurements reported at higher frequencies i.e. PSRs J0206$-$4028, J0828$-$3417, and J1900$-$2600.

\begin{table*}
\centering
\begin{tabular}{l r r r r r r r r r}
\hline \hline
Pulsar   & RA              & Dec           & Ang. Sep.   & $V_{200}$ & SNR & $S_{200}$ (ref) & $v_{200}$ & $v_{\nu}$ & $\nu$ (ref) \\ 
         & (J2000)         &  (J2000)      & (arcmin) & (mJy)     &     & (mJy)            &     (\%)   &   (\%)    & (MHz) \\ 
\hline
J0034$-$0534 & 00$\rah$34$\ram$25$\ras$ & -05$\arcdeg$34$\arcmin$52$\arcsec$ & 0.8 &  $-11.4$ &  5.7 &    65.0 (M) & $-17.5$ &  $-9.8$ &  149 (N)  \\
J0206$-$4028 & 02$\rah$06$\ram$00$\ras$ & -40$\arcdeg$27$\arcmin$19$\arcsec$ & 0.8 &   $-6.2$ &  4.7 &    32.0 (M) & $-19.4$ &  $+9.3$ & 1400 (J)  \\
J0452$-$1759 & 04$\rah$52$\ram$35$\ras$ & -17$\arcdeg$59$\arcmin$08$\arcsec$ & 0.4 &   $+9.1$ &  4.8 &    96.0 (M) &  $+9.5$ &  $+3.6$ & 1400 (J)  \\
J0820$-$4114 & 08$\rah$20$\ram$14$\ras$ & -41$\arcdeg$14$\arcmin$21$\arcsec$ & 0.4 &  $+12.3$ &  4.8 &   116.0 (M) & $+10.6$ &  $+4.5$ & 1400 (J)  \\
J0828$-$3417 & 08$\rah$28$\ram$18$\ras$ & -34$\arcdeg$16$\arcmin$22$\arcsec$ & 0.8 &  $-12.2$ &  4.8 &   400.0 (M) &  $-3.0$ &  $+5.0$ &  606 (Y)  \\
J1239$+$2453 & 12$\rah$39$\ram$39$\ras$ & +24$\arcdeg$53$\arcmin$34$\arcsec$ & 0.4 &  $-54.8$ &  4.3 &   154.7 (B) & $-35.4$ &  $-7.5$ &  149 (N)  \\
J1359$-$6038 & 14$\rah$00$\ram$08$\ras$ & -60$\arcdeg$37$\arcmin$23$\arcsec$ & 1.5 &  $+29.9$ &  5.3 &   402.0 (M) &  $+7.4$ & $+17.3$ & 1400 (J)  \\
J1456$-$6843 & 14$\rah$56$\ram$08$\ras$ & -68$\arcdeg$43$\arcmin$24$\arcsec$ & 0.8 &  $+24.8$ &  5.8 &   738.0 (M) &  $+3.4$ &  $+4.6$ & 1400 (J)  \\
J1543$+$0929 & 15$\rah$43$\ram$40$\ras$ & +09$\arcdeg$28$\arcmin$31$\arcsec$ & 0.8 &  $-29.6$ &  4.8 &   234.0 (M) & $-12.7$ & $-33.0$ &  234 (Y)  \\
J1600$-$5044 & 16$\rah$00$\ram$55$\ras$ & -50$\arcdeg$44$\arcmin$06$\arcsec$ & 0.4 &  $+23.1$ &  5.0 &   139.3 (F) & $+16.6$ & $+29.3$ & 1400 (J)  \\
J1707$-$4053 & 17$\rah$07$\ram$23$\ras$ & -40$\arcdeg$54$\arcmin$11$\arcsec$ & 0.4 &  $-42.6$ &  5.9 &     493 (L) &  $-8.6$ &  $-3.5$ & 1400 (J)  \\
J1834$-$0731 & 18$\rah$34$\ram$17$\ras$ & -07$\arcdeg$31$\arcmin$52$\arcsec$ & 0.8 &  $+30.5$ &  4.2 &             &         & $+14.3$ & 1400 (J)  \\
J1835$-$0643 & 18$\rah$35$\ram$07$\ras$ & -06$\arcdeg$43$\arcmin$21$\arcsec$ & 0.4 &  $-32.8$ &  4.1 &    99.5 (F) & $-32.9$ &  $-5.3$ & 1400 (J)  \\
J1842$-$0612 & 18$\rah$42$\ram$46$\ras$ & -06$\arcdeg$12$\arcmin$51$\arcsec$ & 0.8 &  $+29.1$ &  4.4 &             &         &          &           \\
J1900$-$2600 & 19$\rah$00$\ram$46$\ras$ & -26$\arcdeg$00$\arcmin$59$\arcsec$ & 0.4 &  $-14.4$ &  4.3 &   299.0 (M) &  $-4.8$ &  $+1.4$ & 1400 (J)  \\
J1921$+$2153 & 19$\rah$21$\ram$46$\ras$ & +21$\arcdeg$52$\arcmin$47$\arcsec$ & 0.4 &  $+63.1$ &  5.4 &     914 (L) &  $+6.9$ &  $+6.8$ &  149 (N)  \\
J2241$-$5236 & 22$\rah$41$\ram$37$\ras$ & -52$\arcdeg$35$\arcmin$51$\arcsec$ & 1.1 &  $-12.1$ &  4.7 &    60.0 (M) & $-20.1$ &         &           \\
J2256$-$1024 & 22$\rah$56$\ram$55$\ras$ & -10$\arcdeg$24$\arcmin$49$\arcsec$ & 0.4 &  $+10.0$ &  5.4 &             & $+41.1$ &         &           \\ [1ex]
\hline
\end{tabular}
\caption{Targeted pulsars detected above 4$\sigma$ at 200\,MHz in circular polarisation. Table columns are the same as defined in Table \ref{tab:candidates}. References provided within parenthesis refer to B:\citet{Bilous:2016}, F:\citet{Frail:2016}, J:\citet{Johnston:2017}, L:This work, M:\citet{Murphy:2017}, N:\citet{Noutsos:2015}, and Y:\citet{You:2006}.}
\label{tab:pulsar}
\end{table*}

The pulsar catalogue produced by \citet{Murphy:2017} for all 60 pulsars detected in total intensity at 200 MHz with the MWA provides an accurate sample against which fractional circular polarisation detections and limits can be compared. \citet{Murphy:2017} pulsars that were detected in the blind survey and targeted survey have already been listed in Table \ref{tab:candidates} and Table \ref{tab:pulsar}, respectively. Table \ref{tab:pulsarm} lists all \citet{Murphy:2017} pulsars that were not detected above $4\sigma$ in circular polarisation. When compared against the measured fractional circular polarisation in literature only two of these pulsars, PSRs J0837$-$4135 and J1752$-$2806, were expected to be detected at 200\,MHz above a $4\sigma$ limit. As the previous observations were at 1.4\,GHz, it is likely that the polarimetric behviour of these sources are different at 200\,MHz. In total, 21 out of 60 \citet{Murphy:2017} pulsars are detected above $4\sigma$, a proportion of $35\%$. Additionally, 11 sources were detected in this survey that were not detected by \citet{Murphy:2017}, suggesting that searching for pulsars in circular polarisation can help to discover sources that would have been missed in total intensity searches. 

\begin{table}
\centering
\begin{tabular}{l r r r r r r}
\hline \hline
Pulsar    & $\lvert V_{200}\rvert$ &  $S_{200}$ & $\lvert v_{200}\rvert$ & $v_{\nu}$ & $\nu$ (ref) \\ 
         & (mJy)     &  (mJy)            &     (\%)   &   (\%)    & (MHz) \\ 
\hline
J0737$-$3039A &  $<4.3$ &    53.0 & $<8.1$  &       &       \\
J0809$-$4753  & $<11.8$ &   229.0 & $<5.2$  &  $-0.4$ & 1400 (J) \\
J0820$-$1350  &  $<7.4$ &   160.0 & $<4.6$  &  $-4.2$ & 1400 (J) \\
J0826$+$2637  & $<31.8$ &   243.0 & $<13.1$ &   $+6.2$ & 149  (N) \\
J0837$+$0610  & $<12.4$ &   286.0 & $<4.3$  &  $-2.6$ & 149  (N) \\
J0837$-$4135  & $<11.8$ &    95.0 & $<12.5$ &  $+13.8$ & 1400 (J) \\ 
J0840$-$5332  & $<12.0$ &    56.0 & $<21.4$ &   $+9.0$ & 660  (Y) \\
J0855$-$3331  &  $<9.7$ &    47.0 & $<20.6$ &       &       \\
J0856$-$6137  & $<12.1$ &    85.0 & $<14.3$ &   $+5.6$ & 1400 (J) \\
J0905$-$5127  & $<13.6$ &    73.0 & $<18.6$ &  $+14.1$ & 1400 (J) \\
J0907$-$5157  & $<12.2$ &   106.0 & $<11.6$ &   $+4.8$ & 1400 (J) \\
J0922$+$0638  & $<15.2$ &   100.0 & $<15.2$ &   $+5.6$ & 1400 (J) \\
J0924$-$5302  & $<10.9$ &    96.0 & $<11.3$ &  $-7.3$ & 1400 (J) \\
J0942$-$5552  & $<11.4$ &    73.0 & $<15.7$ &   $+0.6$ & 1400 (J) \\
J0942$-$5657  & $<14.1$ &   112.0 & $<12.6$ &  $+11.6$ & 1400 (J) \\
J0953$+$0755  & $<13.6$ &  1072.0 & $<1.3$  & $-11.5$ & 149  (N) \\
J0959$-$4809  & $<13.6$ &    50.0 & $<27.1$ &  $-4.0$ & 1400 (J) \\
J1001$-$5507  & $<14.3$ &   142.0 & $<10.0$ &   $+1.9$ & 1400 (J) \\
J1012$-$2337  &  $<8.2$ &    47.0 & $<17.4$ &       &       \\
J1047$-$3032  &  $<7.6$ &    24.0 & $<31.8$ &       &       \\
J1057$-$5226  & $<16.2$ &   202.0 & $<8.0$  &   $+3.2$ & 1400 (J) \\
J1116$-$4122  &  $<9.2$ &    52.0 & $<17.6$ &  $-3.6$ & 1400 (J) \\
J1121$-$5444  & $<17.8$ &   101.0 & $<17.6$ &  $-7.8$ & 1400 (J) \\
J1430$-$6623  & $<15.8$ &   190.0 & $<8.3$  &   $+4.5$ & 1400 (J) \\
J1543$-$0620  & $<12.9$ &    91.0 & $<14.1$ &  $-5.0$ & 234  (Y) \\
J1607$-$0032  & $<71.8$ &   137.0 & $<52.4$ &   $+1.4$ & 1400 (J) \\
J1643$-$1224  & $<21.4$ &   123.0 & $<17.4$ &  $-1.0$ & 1331 (Y) \\
J1645$-$0317  & $<23.0$ &   774.0 & $<3.0$  &  $-0.1$ & 1400 (J) \\
J1651$-$1709  & $<20.2$ &   111.0 & $<18.2$ &       &       \\
J1722$-$3207  & $<16.7$ &   229.0 & $<7.3$  &   $+3.9$ & 1400 (J) \\
J1731$-$4744  & $<24.6$ &   325.0 & $<7.6$  &   $+5.4$ & 1400 (J) \\
J1752$-$2806  & $<22.8$ &  1504.0 & $<1.5$  &   $+5.9$ & 1400 (J) \\ 
J1810$+$1744  & $<110.0$ &  231.0 & $<47.6$ &       &       \\
J1820$-$0427  & $<28.9$ &   499.0 & $<5.8$  &  $-3.3$ & 1400 (J) \\
J1824$-$1945  & $<24.3$ &   177.0 & $<13.7$ &   $+1.3$ & 1400 (J) \\
J1824$-$2452A & $<20.0$ &   199.0 & $<10.1$ &       &       \\
J1913$-$0440  & $<20.6$ &   176.0 & $<11.7$ &  $-7.0$ & 149  (N) \\
J2053$-$7200  & $<15.4$ &   110.0 & $<14.0$ &  $-4.0$ & 660  (Y) \\
J2155$-$3118  &  $<7.6$ &    46.0 & $<16.5$ & $-13.8$ & 1400 (J) \\
\hline
\end{tabular}
\caption{Non-detections from \citet{Murphy:2017} catalogue of 200 MHz pulsars. Upper limits are specified at 4$\sigma$ at 200\,MHz in circular polarisation. $S_{200}$ is the total intensity at 200\,MHz taken from \citet{Murphy:2017}. $\lvert v_{200}\rvert$ is the upper limit of the fractional circular polarisation at 200 MHz. Table columns are the same as defined in Table \ref{tab:candidates}. References provided within parenthesis refer to B:\citet{Bilous:2016}, F:\citet{Frail:2016}, J:\citet{Johnston:2017}, L:This work, M:\citet{Murphy:2017}, N:\citet{Noutsos:2015}, and Y:\citet{You:2006}.}
\label{tab:pulsarm}
\end{table}

We note that the pulsar PSR J2330$-$2005, previously detected by \citet{Lenc:2016} in deep observations at 154 MHz, is not detected in the targeted survey. The source was found to be circularly polarised with a flux density of $-8.9\pm1.1$\,mJy and $-9.6\pm1.0$\,mJy in two separate epochs at 154\,MHz. In the 200 MHz all-sky survey, our $4\sigma$ limit is $\lvert V_{200}\rvert<7.2$ \,mJy PSF$^{-1}$ for this source location. When adjusted for the GLEAM spectral index of this source ($\alpha=-0.71\pm0.57$), the brightest detection of this source would have an expected circular polarisation of $-8.0\pm1.6$ at 200\,MHz ($-7.4\pm1.6$ for the weaker detection). Given the error constraints, it is possible that this source may have fallen below the threshold of this survey. Deeper observations would be required to confirm the nature of this source at 200\,MHz. 

\subsection{Limits on radio emission from exoplanets}\label{sec:exoplanet}

The magnetised planets in our Solar system emit intense, low-frequency radio emission associated with planetary aurora. Similarly, planets outside our Solar System (i.e. exoplanets) capable of generating planetary-scale magnetic fields are expected to produce bright radio emission \citep{Winglee:1986, Zarka:2001}. The emission is produced via the electron-cyclotron maser (CMI) instability, arising from the propagation of energetic electrons along converging magnetic field lines in the magnetosphere of the planet. CMI emission is characterised as bright, beamed, highly circularly polarised radio emission that can be variable on time-scales of seconds to days \citep{Wu:1979, Treumann:2006}. Because the emitting frequency of the planetary radio emission is tied magnetic field strength, radio detections of exoplanets will directly measure their field strengths and in turn provide insight into the interior composition of these planets. Additionally, the variations of the radio emission in time and frequency variability can provide geometrical constraints on the planet's orbit and magnetic field configuration \citep{Hess:2011}.

There have been many observational attempts to detect radio emission from exoplanets \citep{Bastian:2000, Lazio:2004, Lazio:2007, George:2007, Smith:2009, Lazio:2010, Stroe:2012, Lecavelier:2013, Hallinan:2013, Sirothia:2014, Murphy:2015, Lynch:2017a, OGorman:2018} but there have been no unambiguous detections to date. The expected high fractional circular polarisation of CMI emission makes exoplanets prime targets for Stokes V searches. Two previous studies have used  Stokes V imaging with the MWA to search for radio emission from exoplanets. From a catalog of 1\,110 known exoplanets (as of 2014 May 14), \citet{Murphy:2015} targeted 17 sources that they identified as having estimated flux densities and emission frequencies close to or above the MWA detection capabilities. \citet{Lynch:2017a} observed a young star forming region to search for variable Stokes V emission that might be associated with exoplanets in still forming planetary systems. 

Since the publication of \citet{Murphy:2015} many thousands of exoplanets have been discovered through various optical techniques \citep{Schneider:2011}. Using an updated catalog of 4\,132 sources (known population of exoplanets as of 2018 February 19), we did a targeted search of the 1\,506 sources located within our survey region for significant circularly polarised emission above the estimated local image noise. Again, we used a lower, 4$\sigma$ threshold since an a priori position was known. Of the 1\,506 sources searched, two sources, Proxima Cen b and HD 34445 b, were found to be associated with emission at a $>$4$\sigma$ level. 

Visually inspecting the Stokes V image for HD 34445 b, we found the source have structure in the image that was indicative of a noise peak in the image; thus we ruled this source out as a detection. Visual inspection of the Proxima Centauri b image found the emission to be point-like and an investigation of the associated Stokes I emission did not reveal a bright source that could be responsible for any total intensity leakage. We tentatively claim to make a detection of weak emission at the location of Proxima Centauri b, however the detected radio emission is not expected to be associated with the planet but instead with the host star. 

CMI emission is emitted at the cyclotron frequency of the source population of the electrons, which is directly related to the local magnetic field strength, $B_p$, of the planetary magnetosphere:
\begin{equation}
f_c = \frac{eB_p}{2\pi m_e}\approx 2.8\ \text{MHz}\ B_p
\end{equation}
where, $m_e$ and $e$ are the electron mass and charge, and $B_p$ is measured in Gauss. The maximum estimated magnetic field strength for Proxima Centauri b is 1\,Gauss \citep{Burkhart:2017}, corresponding to a maximum emission frequency of $\sim$3 MHz. Due to ionospheric absorption of emission at frequencies $<$10\,MHz, planetary radio emission from Proxima Centauri b cannot be detected by ground-based radio telescopes. Thus any emission that we detected using ground based radio telescopes must be related to the magnetic activity of the star; the possibility of Proxima Centauri producing the observed emission is discussed the next section. 

The upper limits set by this survey for a set of best radio detection candidate exoplanets, as identified by their theoretically estimated emission frequencies and radio flux densities, will be discussed in a future paper (Lynch et al. \emph{in prep}). 

\subsection{Limits on radio emission from flare stars}\label{sec:flareStars}

Some magnetically active stars are observed to exhibit short-duration, narrow band, and highly circularly polarised ($\sim$100$\%$) radio flares. The observed polarisation and frequency-time structure of these flares points to a coherent emission mechanism such as CMI \citep{Bastian:2000, Gudel:2002}. In the 1960s -- 1970s several magnetically active M dwarf stars were observed at frequencies between $90 - 300$~MHz using single dish telescopes. These observations revealed bright radio flares with rates between $0.03 - 0.8$ flares per hour and intensities ranging from $0.8$ to $20$\,Jy. However, recent low-frequency surveys to detect transients have resulted in non-detections \citep[e.g.][]{Rowlinson:2016, Tingay:2016}. 

To confirm the previous M dwarf stellar flare rates and flux densities at $100 - 200$\,MHz, \citet{Lynch:2017b} targeted UV Ceti, a magnetically active M dwarf star. As the radio flares from UV Ceti were expected to be highly circularly polarised, this search was focused in the circularly polarised images rather than in total intensity. Four flares were detected from UV Ceti with flux densities a factor of 100 fainter than those in the literature. 

Following this example we used the updated catalog of radio stars by \citet{Wendker:2015} to search for circularly polarised emission associated with the positions of these objects. A wide variety of stellar objects are included in this catalogue including M dwarf stars, RS CVn binaries, and magnetic chemically peculiar hot stars. This catalog contains 3\,021 objects, 2\,400 of which are located within our survey region. From this search we identify 3 objects associated with emission at a $>$4$\sigma$ level: Proxima Centauri, HR 5942, and DM-64 1685. Visual inspection of the circularly polarised image for D-64 1685 ruled it out as a detection, as the emission structure more closely resembles that of image noise than point-source like. In the total intensity image, the location of HR 5942 offset from a bright extragalactic source leaving us to conclude that the observed Stokes V emission is not due to total intensity leakage. A similar offset is found for both Stokes Q and U and not indicative of linear polarisation leakage. We tentatively claim a $4.5\sigma$ detection of HR 5942 with a measured Stokes V flux density of $-11\pm 3$\,mJy. Additionally, we claim a tentative detection of Proxima Centauri, with a measured flux density of $-18\pm 4$\,mJy.

Both HR 5942 and Proxima Centauri are previously detected in the radio. HR 5942 is a magnetic chemically peculiar Bp star with previous detections of quiescent emission at 5 and 14\,GHz \citep{Linsky:1992, Leone:1994} and radio flaring at 5\,GHz \citep{Drake:1989}; both types of emission are thought to be gyrosynchrotron emission. Coherent emission has been observed in other magnetic chemically peculiar stars from 610 to 1400\,MHz \citep{Trigilio:2000, Chandra:2015, Das:2018}. If the detection of HR 5942 is confirmed, this would be the lowest frequency detection of a magnetic chemically peculiar hot star and only the third hot star to be observed emitting highly circularly polarised, coherent emision. Proxima Centauri is an emission-line M dwarf star, previously observed to emit bright coherent bursts at $\sim$1\,GHz \citep{Lim:1996, Slee:2003}. Other M dwarf stars have been observed to emit radio emission from MHz to GHz frequencies (e.g. AD Leo: \citet{Spangler:1974}, \citet{Jackson:1989}; YZ CMi: \citet{Spangler:1976}, \citet{Kundu:1988}). Given the previously observed GHz bursts from Proxima Centauri, it is possible that this source could also produce bursts at 170 -- 230\, MHz; a previous 3$\sigma$ limit of 42.3\,mJy at 200\,MHz has been reported by \citet{Bell:2016b}. To further confirm the tentative detections of HR 5942 and Proxima Centauri, investigations into the variability and frequency spectrum of the observed emission is ongoing. 

\section{Conclusions}

We have demonstrated the effectiveness of polarisation leakage mitigation using MWA observations and have used it to complete an all-sky survey in circular polarisation using existing observations. The fractional leakage was typically reduced by an order of magnitude to less than $0.72\%$ and allowed both blind and targeted surveys to be performed with a sensitivity of $1.0-3.0$\,mJy\,PSF$^{-1}$.

We have detected 32 pulsars, 6 transient emissions from artificial satellites and 2 flare stars. When compared against total intensity observations of pulsars at 200\,MHz, $35\%$ of pulsars that were detectable in total intensity were also detected in circular polarisation. Furthermore, 11 pulsars detected in circular polarisation were not originally found in total intensity (as a result of their location in the Galactic plane or the limited sensitivity available in Stokes I because of confusion). The 2 flare stars detected in this survey were only detected in circular polarisation due to either limited sensitivity in the total intensity image or the close proximity of a nearby, bright extragalactic source. Of the 3\,610 exoplanets in our catalogue of known objects, 1\,506 exoplanets were located within our survey region; these were also searched but did not yield any detections.

The all-sky survey presented here was not ideal for detecting transient emission from sources such as flare stars and exoplanets. Transient sources and sources that can change sign in polarisation, such as seen with the flare star UV Ceti \citep{Lynch:2017b}, require an alternate observing and processing strategy. To avoid diluting the signal, the integration of the snapshot images should not exceed the time-scale of the expected emission before a sign flip occurs or the emission stops. Similarly, for periodic emission where the duty cycle is low, tracked observations of a field would be better suited to increase the probability of catching the moment of the flaring emission. Two avenues of investigation what will be pursued in future will be to search through overlapping snapshot images of the drift-scan for transient emission and to apply the leakage mitigation techniques developed here to targeted observations.

While an order-of-magnitude improvement in Stokes I to Stokes V leakage has been greatly beneficial, further improvements would be required to probe sources with low levels of fractional circular polarisation. The technique presented in this paper is currently limited by sidelobe confusion, noise and fitting of the 2D model of the position-dependent leakage. A more extended antenna array, such as that available with the recent MWA extension, can help reduce PSF sidelobes and improve the sensitivity of uniform-like imaging. Sensitivity can also be improved by avoiding beam-former or frequency changes over the course of a drift-scan. With a near-continuous drift-scan, field sources can probe leakage over much finer tracks throughout the field. Finally, improved 2D modelling can also help to reduce errors at the field edges where increased residual leakage is noticed. Currently a simply 2D quadratic function is used for fitting: more complex functions may improve the fitting results.

An outstanding challenge not addressed in the survey presented here is distinguishing between Stokes U to Stokes V leakage and true circular polarisation. This is only problematic for sources with both a significant degree of linear polarisation and with a low rotation measure. As such, it only affects a small sub-set of candidate sources. To determine the effect of this leakage requires knowledge of the X-Y phase which is typically obtained by observing a linearly polarised source. Such sources are rare at long wavelengths, however, it may be possible to measure the effect in diffuse linearly polarised Galactic emission \citep{Lenc:2017}. The practicality and effectiveness of this is yet to be investigated for an all-sky survey.

The methods for leakage mitigation demonstrated here should also be applicable to Square Kilometre Array Low Frequency array (SKA-LOW \footnote{See SKA phase 1 system (Level 1) requirements SKA-TEL-SKO-0000008: \url{http://www.skatelescope.org/wp-content/uploads/2014/03/SKA-TEL-SKO-0000008-AG-REQ-SRS-Rev06-SKA1_Level_1_System_Requirement_Specification-P1-signed.pdf}}. The method requires minimal processing and is fast because no deconvolution is required. The main requirement is that the nature of the leakage remains constant for instrumental beam. A secondary requirement is that good quality images can be generated on relatively small time-scales. In the case of the MWA, the large number of available baselines enables it to generate good quality images on 2-minute time-scales. The current specification for SKA-low has fewer baselines compared to the MWA but they will be more sensitive and more extended. As long as this compromise does not adversely affect the quality of the snapshot dirty maps, the mitigation techniques used for the MWA should still be effective with SKA-Low.

\section*{Acknowledgments}
The authors thank Ron Ekers for useful discussions. TM acknowledges the support of the Australian Research Council through grant FT150100099. DLK was supported by NSF grant AST-1412421. This scientific work makes use of the Murchison Radio-astronomy Observatory, operated by CSIRO. We acknowledge the Wajarri Yamatji people as the traditional owners of the Observatory site. Support for the operation of the MWA is provided by the Australian Government (NCRIS), under a contract to Curtin University administered by Astronomy Australia Limited. We acknowledge the Pawsey Supercomputing Centre which is supported by the Western Australian and Australian Governments. This research was conducted by the Australian Research Council Centre of Excellence for All-sky Astrophysics (CAASTRO), through project number CE110001020. This research has made use of the NASA/IPAC Extragalactic Database (NED) which is operated by the Jet Propulsion Laboratory, California Institute of Technology, under contract with the National Aeronautics and Space Administration. The authors thank the anonymous referee for providing useful comments on the original version of this paper.

\bibliographystyle{mnras}
\bibliography{biblio} 

\end{document}